\documentstyle[12pt]{article}
\input epsf.tex

\newcommand{\bmat}{\left(\begin{array}}
\newcommand{\emat}{\end{array}\right)}

\def\yzero{\smash{\hbox{$y\kern-4pt\raise1pt\hbox{${}^\circ$}$}}}

\def\-{\hphantom{-}}
\def\ov{\overline}
\def\s2{\frac{1}{\sqrt2}}

\def\beq{\begin{equation}}
\def\eeq{\end{equation}}
\def\beqa{\begin{eqnarray}}
\def\eeqa{\end{eqnarray}}
\def\tr{{\rm tr \,}}

\def\diag{{\rm diag \,}}

\def\IF{\relax{\rm I\kern-.18em F}}
\def\II{\relax{\rm I\kern-.18em I}}
\def\IP{\relax{\rm I\kern-.18em P}}

\def\NN{{\cal N}}

\def\Dsl{\,\raise.15ex\hbox{/}\mkern-13.5mu D} %can be subscripted

\def\IC{\bf C}
\def\IZ{\bf Z}
\def\IS{\bf S}
\def\z2z2{$\IC^3/(\IZ_2\times\IZ_2)$}

\def\id{{\bf 1}}

\newcommand{\drawsquare}[2]{\hbox{%
\rule{#2pt}{#1pt}\hskip-#2pt%  left vertical
\rule{#1pt}{#2pt}\hskip-#1pt%  lower horizontal
\rule[#1pt]{#1pt}{#2pt}}\rule[#1pt]{#2pt}{#2pt}\hskip-#2pt%  upper horizontal
\rule{#2pt}{#1pt}}% right vertical

\newcommand{\fund}{\raisebox{-.5pt}{\drawsquare{6.5}{0.4}}}%  fund
\newcommand{\antifund}{\overline{\fund}}

\topmargin
-1.5cm
\textwidth
15.5cm
\textheight
24cm
\oddsidemargin
0.7cm
\evensidemargin
1.2cm

\begin{document}

%----------------------------------------------------------------------%
%  numbering equations with section number
%----------------------------------------------------------------------%
\makeatletter
\@addtoreset{equation}{section}
\makeatother
\renewcommand{\theequation}{\thesection.\arabic{equation}}
%----------------------------------------------------------------------%
%  title page
%----------------------------------------------------------------------%
\pagestyle{empty}
%\vspace*{1.0in}
\rightline{IFT-UAM/CSIC-05-10}

\rightline{CERN-TH-PH/2005-038}

\rightline{\tt hep-th/0503079}
\vspace{0.5cm}
\begin{center}
\LARGE{\bf Holographic dual\\ of the Standard Model on the throat}

\vspace{0.5cm}

\large{
J. F. G. Cascales$^1$, F. Saad$^1$, A.~M.~Uranga$^{1,2}$} \\[4mm]
$^1$ {\em Instituto de F\'{\i}sica Te\'orica, Facultad de Ciencias}\\
{\em Universidad Aut\'onoma de Madrid, 28049 Madrid, Spain}\\[4mm]
$^2$ {\em Theory Division, CERN}\\
{\em CH-1211 Geneva 23, Switzerland} \\[4mm]

\vspace*{2cm}

\small{\bf Abstract} \\[7mm]
\end{center}

\begin{center} \begin{minipage}[h]{14.0cm}
{\small We apply recent techniques to construct geometries, based on local 
Calabi-Yau manifolds, leading to warped throats with 3-form fluxes in 
string theory, with interesting structure at their bottom. We provide
their holographic dual description in terms of RG flows for gauge theories 
with almost conformal duality cascades and infrared confinement. We 
describe a model of a throat with D-branes at its bottom, realizing a 
3-family Standard Model like chiral sector. We provide the explicit 
holographic dual gauge theory RG flow, and describe the appearance of the 
SM degrees of freedom after confinement. As a second application, we 
describe throats within throats, namely warped throats with discontinuous 
warp factor in different regions of the radial coordinate, and discuss 
possible model building applications.}

\end{minipage}
\end{center}

\newpage
%----------------------------------------------------------------------%
%  Resetting of counters
%----------------------------------------------------------------------%
\setcounter{page}{1}
\pagestyle{plain}
\renewcommand{\thefootnote}{\arabic{footnote}}
\setcounter{footnote}{0}
%----------------------------------------------------------------------%
%  Paper begins
%----------------------------------------------------------------------%

\section{Introduction}

Theories with strongly warped extra dimensions (warped throats) have 
revealed novel features compared with standard factorised 
compactifications \cite{rs1,rs2}. Such theories have been intensively 
applied to phenomenological model building beyond the Standard Model (SM) to address 
a variety of questions. The prototypical example of such applications is 
the RS1 construction \cite{rs1}, where a slice of $AdS_5$, with two 
boundaries, is regarded 
as a 5d compactification to 4d with an exponential warp factor in the 
extra dimension. Location of SM fields at the strongly warped end 
(infrared brane) leads to an exponential suppression of 4d scales as 
compared with the mildly warped end (ultraviolet brane), thus providing a 
new approach to the Planck/electroweak hierarchy. Inspired by the AdS/CFT 
correspondence in string theory, results in RS phenomenology have been 
interpreted in purely 4d terms by replacing the warped 
throat by a strongly interacting 4d conformal field theory. 
However, lack of a microscopic understanding of holography in the 
effective field theory approach prevents this picture to go beyond a 
qualitative rephrasing.

A microscopic construction of warped throats and their holographic 
description can be obtained in string theory. Warped extra dimensions 
appear in string theory in compactifications with non-trivial field 
strength fluxes, due to their backreaction on the underlying metric
\cite{drs,gkp}. In particular, warped throats with exponential warp 
factors appear when fluxes are (in intuitive terms) associated to 3-cycles 
localized in small regions of the compactification space. 

The prototypical example is provided by the Klebanov-Strassler throat 
\cite{ks}. It is based on the local geometry of the deformed conifold. 
One turns on $M$ units of RR 3-form flux on the $\IS^3$ at its tip, and 
a suitable density of NSNS 3-form flux on its dual non-compact 3-cycle. 
If the latter is considered compact by setting a cutoff distance (or by 
embedding in a global compactification), one denotes by $K$ the total NSNS 
flux. For $K\gg g_sM$, where  $g_s$ is the IIB string coupling, fluxes 
backreact on the geometry and create a warped throat, whose geometry is 
approximately $AdS_5\times T^{1,1}$ (with the cosmological constant of 
$AdS_5$ and the 5-form over $T^{1,1}$ slowly varying along the radial 
direction). The roles of the IR and UV branes in RS1 are played by the 
$\IS^3$ of the deformed conifold and the cutoff/compactification, 
respectively. The relative warp factors between both endpoints is 
$e^{-\frac{2\pi K}{3g_sM}}$ \cite{verlinde,ks,gkp}.
The holographic dual description is given by 
the (almost conformal) $\NN=1$ supersymmetric 4d gauge theory on $N=KM$ 
D3-branes at the singular 
conifold in the presence of $M$ fractional branes. The gauge theory 
\cite{kw} has a gauge group $SU(N)\times SU(N+M)$, chiral multiplets 
$A_i$, $B_i$, $i=1,2$ in the representations $(\fund,\antifund)$, 
$(\antifund,\fund)$ respectively, and a superpotential $W= 
\epsilon^{ij}\epsilon^{kl} \tr A_i B_k A_j B_l$, which is marginal for the 
strongly interacting conformal theory. Along the renormalization group 
(RG) flow to the infrared, the theory suffers a cascade of Seiberg 
dualities \cite{seiberg}, in which the effective $N$ decreases in steps 
of $M$. Eventually, at an infrared scale achieved after $K=N/M$ duality 
steps, the theory confines and running stops. This corresponds to the IR 
ending of the dual throat, with the infrared confinement scale related to the 
$\IS^3$ size. The warp factor is associated to the ratio of UV and IR 
scales $e^{-\frac{2\pi K}{3 g_s M}}$ generated by the RG flow.

Unfortunately, the Klebanov-Strassler throat is too simple to e.g. 
generate chiral physics in the infrared. On the gravity side, the geometry is 
smooth after the deformation, while in the field theory side, the light 
degrees of freedom are simply the glueballs of the confining theory. Hence
the throat does not allow embedding the SM degrees of freedom at the IR 
end. An attempt to construct throats with more structure at their bottom 
was carried out in \cite{cgqu}, but involved complicated geometries whose 
holographic dual was really unknown.

Happily, new progress in understanding warped throats for other 
geometries generalising the conifold, as well as their 
interpretation in terms of duality cascades \cite{chaotic,ejaz,fhu} and 
infrared confinement \cite{fhu}, allows to revisit these ideas. 
The general lesson is that warped throats arise naturally by 
considering Calabi-Yau singularities which admit a complex deformation, 
corresponding to smoothing the singular point by growing a set of 
3-cycles, on which to turn on fluxes. The holographic duals are given by 
duality cascades of the gauge theories arising on D3-branes at the singular 
geometries, much in the spirit of Vafa's brane-flux transitions 
\cite{vafa}. 

\begin{figure}[ht]
  \epsfxsize = 8cm
  \centerline{\epsfbox{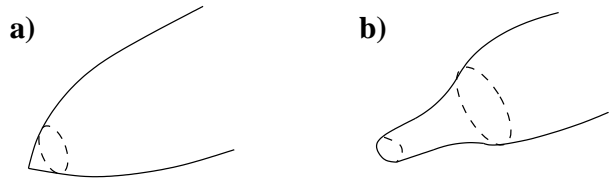}}
  \caption{\small Throats with two different infrared behaviours: Figure 
a) shows a throat based on a geometry which upon complex deformation 
leaves a `terminal' singularity. Figure b) shows a throat based on a 
geometry which upon complex deformation leaves a geometry admitting a 
further complex deformation (in this case to a completely smooth space). }
  \label{twopos}
\end{figure}

This allows the description of warped throats, and their holographic 
duals, with richer infrared topology/dynamics. This paper is devoted 
to exploring some of the new model building possibilities of these 
constructions. In particular, an amusing feature is that the geometries 
after complex deformation may still contain a milder singularity. This is 
depicted in figure \ref{twopos}a. Locating D3-brane probes at such 
singularities corresponds to localising a chiral gauge sector at the IR 
brane of the warped throat construction. In the holographic picture this 
corresponds to a cascading gauge theory which after confinement leaves a 
set of light degrees of freedom whose effective field theory is a chiral 
gauge theory. This motivates our first 
application. We describe an explicit  construction of a warped throat, and 
its holographic dual cascading theory, leading to a chiral gauge 
sector describing a 3-family SM-like chiral gauge theory. In 4d terms, it 
corresponds to a (supersymmetric) walking technicolor model, where all SM 
fields are composites of a confining theory, which is almost conformal in 
the ultraviolet, as we describe explicitly.

A second possibility is that the singularity at the bottom of the throat 
admits a further complex deformation, and hence an additional set of 
3-cycles and fluxes. This corresponds to the development of a new 
throat whose UV is patched with the IR of the initial throat, as shown in 
figure \ref{twopos}b. The general case corresponds to a throat made up of 
several regions in the radial direction, corresponding to different warp 
factors, and separated by  thin shells associated to the appearance of a 
3-cycle supporting some flux \cite{fhu}. In the holographic picture, this 
corresponds to the RG flow of a cascading gauge theory, which experiences 
several scales of partial confinement, after which the cascades of the 
remaining gauge theory follow. This provides a 
string theory description of RS constructions with several positive 
tension branes \cite{oda,severalRS}. Following \cite{fhu} we describe one 
example of this 
kind, and  point out possible applications of such multi-warp throats.

Our models should be regarded as illustrating the general techniques to 
construct throats suitable for model building applications, and their 
holographic dual field theories. Clearly, many other generalisations and 
extensions are possible. We expect much progress in model building 
applications of the new throats, and, although the constructions are 
supersymmetric (in order to automatically guarantee the stability of the 
throat, and to have control over the strongly coupled gauge theory 
dynamics), also in extending similar constructions in non-supersymmetric 
setups.

The paper is organised as follows. In Section \ref{deformed} we review the 
tools to describe singularities admitting complex deformations, in which 
the singularity is replaced by a set of localized 3-cycles. In Section 
\ref{singus} we construct throats with singularities and chiral gauge 
sectors at its IR end, and their holographic duals. After a discussion of 
different interesting possibilities in section \ref{general}, we centre 
on a particular case in section \ref{orbifoldofspp}. We describe the 
throat with a set of D3-branes at a $\IC^3/\IZ_3$ singularity at its 
bottom, and the holographic dual picture. In section \ref{d7branes} we 
improve the model by the addition of D7-branes in the throat picture, 
corresponding to adding flavours to the dual field theory. In section 
\ref{multiwarp} we describe examples of multi-warp throats and possible 
applications. Section \ref{conclusions} contains our final remarks. 
Appendix \ref{appspp} contains results for a local Calabi-Yau singularity 
related to that in section \ref{singus}, and its field theory dual.

\section{Complex deformed spaces}
\label{deformed}

The natural way to construct warped throats with UV and IR cutoffs in 
string theory is via the introduction of 3-form fluxes, localized on a 
set of 3-cycles.
The backreaction of the fluxes creates the warp factor. The 
compactification provides the UV cutoff, while the finite size of the 
3-cycles cuts off the throats in the IR.

A natural way to obtain localized 3-cycles is to consider singular 
geometries and carry out a complex deformation. For instance, the deformed 
conifold arises this way, and so do other throats considered in 
\cite{fhu}. Hence we are led to the study of complex deformations of 
local singularities.

\begin{figure}[ht]
  \epsfxsize = 10cm
  \centerline{\epsfbox{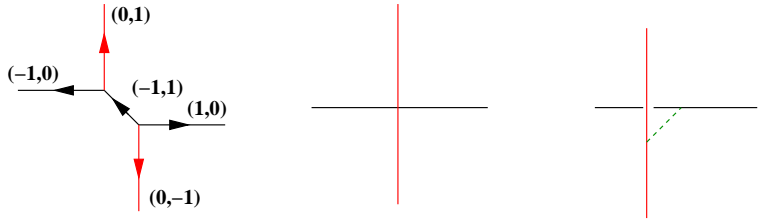}}
  \caption{\small Conifold extremal transition. The finite segment in the 
first figure represents an $S^2$, while the dashed segment in the 
last figure corresponds to an $S^3$.}
  \label{conitrans}
\end{figure}

In many interesting cases, these are simply described in terms of toric 
diagrams. For details we refer the reader to \cite{aganagicvafa,fhu}. For 
our purposes, it suffices to say that the geometries we consider can be 
encoded in web diagrams of segments on a 2-plane, carrying $(p,q)$ 
labels, with the condition that the slope of the segment is related to 
the label, and that there is conservation of the label charge at 
junctions of segments \cite{pqwebs}. The web 
encodes the geometry by specifying the locus on which certain $\IS^1$ 
fibers degenerate \cite{vafaleung,aganagicvafa}. Interior segments and 
faces correspond to finite-size 
2- and 4-cycles. The singular geometries where these shrink to zero size 
sometimes admits complex deformations, where finite-size 3-cycles appear. 
These are represented by a separation (in a new direction) of sub-webs in 
equilibrium. The 3-cycles can be encoded as segments joining the 
different sub-webs. The case of the conifold is shown in figure 
\ref{conitrans}, and other examples will appear in the paper. See 
\cite{fhu} for more examples.

Geometries with finite size 3-cycles can support 3-form fluxes. 
Specifically, one can turn on RR 3-form flux on the compact 3-cycles and 
NSNS 3-form flux on the dual (non-compact) 3-cycles. This leads to warped 
throats of the kind we would like to consider. The relation between the 
size of the 3-cycle and the warp factor with the fundamental parameters 
(the flux quanta) is exponential, leading to a hierarchy. 

The bulk of these throats is approximately $AdS_5$, so they should admit a 
dual holographic description in terms of an approximately conformal field 
theory. Indeed throats arising from geometries using toric diagrams as 
above have a natural holographic interpretation. The dual corresponds to a 
duality cascade in a quiver gauge theory. The quiver gauge theory is given 
by the world-volume theory of D3-branes located at the singularity of the
geometry in the singular limit (where 2/4-cycles and 3-cycles have zero 
size). The amount of flux is encoded in the presence of 
certain fractional branes in the dual, which in the regime of being much 
smaller than the total number of branes, leads to an approximately 
conformal theory. Along the duality cascade the gauge theory suffers a 
cascade of Seiberg dualities, and reduces its number of degrees of freedom. 
At the end of the cascade, there is strong gauge dynamics, confinement 
etc, with a characteristic scale holographically related to the size of 
the 3-cycle. 

When there are several 3-cycles, the fluxes on them are independent. 
Hence one can obtain different sizes for the 3-cycles, with sizes 
hierarchically related to each other. This corresponds to a throat, at the 
bottom of which we trigger one complex deformation and are left over with 
a new, simpler geometry (which may subsequently suffer a further 
complex deformation etc). In the 
holographic picture, we have a field theory which cascades down until a 
strong dynamics scale. Below this scale, we have a new theory, which 
describes the left over geometry after deformation. If this new theory 
continues cascading, it corresponds to a further throat etc \cite{fhu}.

In the coming sections we describe the geometries (and holographic dual 
field theories) illustrating these two behaviours, namely throats 
admitting several complex deformations or throats ending up on singular 
geometries. In section \ref{singus} 
we describe a throat ending in a singularity at which we localise 
D-branes with the SM degrees of freedom. In section \ref{multiwarp} we 
describe throats ending on configurations leading to further throats.

\section{Singularities within throats}
\label{singus}

\subsection{General discussion}
\label{general}

A natural way to realize chiral gauge theories is with D3-branes at 
singularities \cite{dm,dgm}. In fact, there are models reasonably 
similar to the SM with D3- and D7-branes at the $\IC^3/\IZ_3$ singularity 
\cite{aiqu}. 

So in order to obtain a chiral gauge sector from D3-branes at a 
singularity $X$, at the bottom of a warped throat, what we need is a 
geometry $Y$ which admits a complex deformation, leaving $X$ as the left 
over geometry. The 3-cycle associated to the complex deformation supports 
the flux creating the warped throat.

In what follows, we describe a technique to construct spaces $Y$ which 
admit a complex deformation to a given singular space $X$. Although the 
procedure is general for toric varieties, for concreteness we centre on 
the case where the left over geometry $X$ of interest is a $\IC^3/\IZ_3$ 
singularity. Other examples can be worked out similarly.

\begin{figure}[ht]
  \epsfxsize = 8cm
  \centerline{\epsfbox{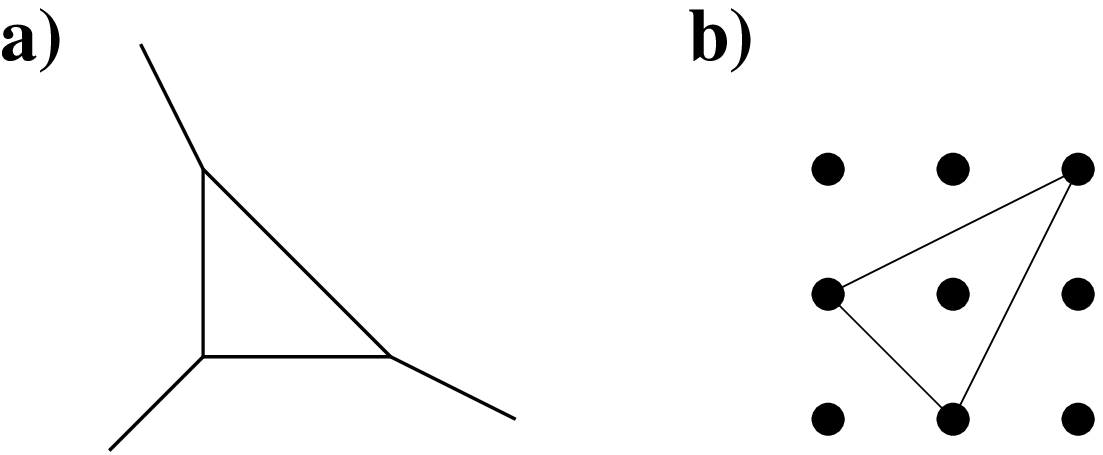}}
  \caption{\small Web diagram and toric diagram for the $\IC^3/\IZ_3$ 
orbifold.}
  \label{z3}
\end{figure}

Hence we consider the geometry $\IC^3/\IZ_3$. This corresponds to 
the web diagram shown in figure \ref{z3}a, and the toric diagram shown 
in figure \ref{z3}b (for toric or grid diagrams, see \cite{pqwebs}).
Locating a set of D3-branes at this 
singular space leads to an $\NN=1$ supersymmetric gauge theory with the 
following structure
\beqa
 {\rm Vect. Mult.}&\quad SU(n)^3 &\nonumber \\
{\rm Ch. Mult.}& \quad 3(\fund,\antifund,1) + 3(1,\fund,\antifund) + 
3(\antifund,1,\fund) &
\eeqa

\begin{figure}[ht]
  \epsfxsize = 12cm
  \centerline{\epsfbox{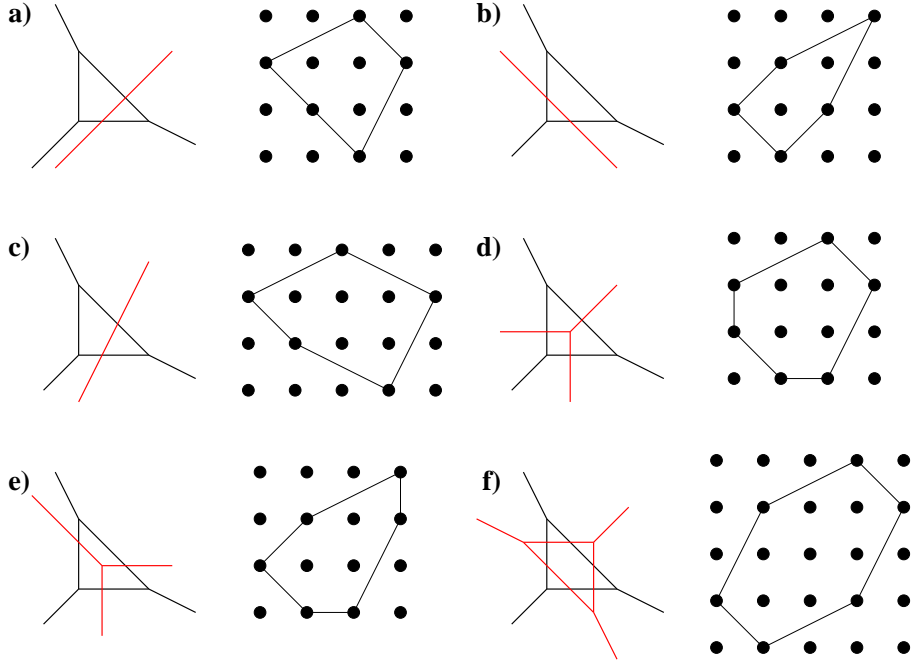}}
  \caption{\small Web and toric diagrams for diverse examples of 
geometries $Y$ which admit a complex deformation to the $\IC^3/\IZ_3$ 
singularity. The web diagrams are shown at the origin of the deformation 
branch, and the subwebs to be removed is shown in red.}
  \label{possibil}
\end{figure}

We would like to consider spaces $Y$, which admit a complex deformation 
to $X$. This search is simpler if we require $Y$ to be toric. Hence we 
need to look for web diagrams, which upon removal of a subweb in 
equilibrium leaves diagram \ref{z3}a behind. Alternatively, webs for $Y$ 
can be constructed by adding webs in equilibrium to figure \ref{z3}a. It 
is important to realize that the final geometry of $Y$ depends only on 
the asymptotic legs added, and not on details like the position of the 
new diagram etc. Some examples of such geometries are shown in figure 
\ref{possibil}.

Now turning on fluxes on the complex deformed space leads to a warped 
throat, cutoff in the infrared by the finite-size 3-cycle, that moreover 
contains a singularity locally of the form $\IC^3 / \IZ_3$ 

It would be very useful to have a holographic description of these 
throats. This requires knowing the quiver gauge theory of D3-branes at the 
space $Y$, in the regime where it is singular. Unfortunately  this 
information is available only for a few classes of singularities. These
include the simplest singularities (conifold, suspended pinch point, etc)
\cite{kw,mp,uraconi}, complex cones over del Pezzo surfaces \cite{delpezzo},
and real cones over the $Y^{p,q}$ 5d horizons studied in \cite{ypq}. Also, 
there are systematic tools to construct (and identify) the quiver field 
theories for orbifold quotients of any of the above geometries (and 
obviously of flat space).

As is illustrated by the examples in figure \ref{possibil}, the geometries 
get more complicated as the structure of the subweb to be removed becomes 
more involved. In general, the quiver gauge theory on D3-branes at such 
singularities is not known. However, there are a few special cases where 
it can be constructed. For instance, figure \ref{possibil}a corresponds to 
a quotient of the suspended pinch point, a singularity for which the 
quiver gauge theory is known. Similarly, figure \ref{possibil}f 
corresponds to a quotient of the cone over $dP_3$. In such cases, one can 
construct the quiver theories for the quotient space from the quiver 
theories for the parent space, in a systematic way. We will 
centre on one such example, to be discussed below. Nevertheless, we would 
like to emphasise that the general ideas are valid, even for geometries 
which are not orbifolds of simpler spaces, once the quiver gauge theories 
are known. In particular, some other geometries in figure \ref{possibil} 
may allow the determination of the quiver gauge theory \cite{private} by 
techniques of unHiggssing \cite{unhiggs}.

\subsection{The orbifold of the SPP}
\label{orbifoldofspp}

\begin{figure}[ht]
  \epsfxsize = 8cm
  \centerline{\epsfbox{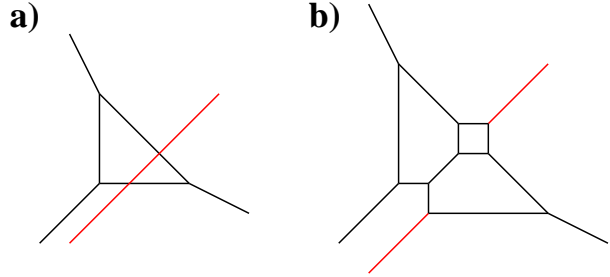}}
  \caption{\small Web diagram for a geometry which admits a complex deformation 
to the $\IC^3/\IZ_3$ orbifold. Figure a) shows the geometry at the origin 
of the deformation branch, while figure b) shows the geometry with finite 
size 2- and 4-cycles.}
  \label{sppz3}
\end{figure}

\begin{figure}[ht]
  \epsfxsize = 8cm
  \centerline{\epsfbox{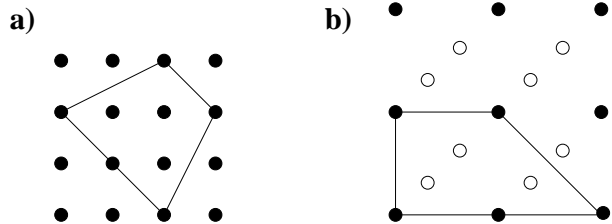}}
  \caption{\small Toric diagram for the geometry in figure \ref{sppz3}. By 
a linear transformation the toric diagram (figure b) manifestly shows that 
the geometry corresponds to a $\IZ_3$ quotient of the suspended pinch point 
singularity. Namely, the black dots represent the toric diagram of the 
SPP geometry, and our geometry is obtained upon a refinement of the toric 
lattice, shown as open circles.}
  \label{toricsppz3}
\end{figure}

Among the deformed spaces corresponding to the pictures in figure 
\ref{possibil}, we centre on that shown in figure \ref{sppz3}a. This is 
because it corresponds to geometry for which the quiver gauge theory is 
easily obtained. Indeed, it corresponds to an orbifold of a singularity, 
the suspended pinch point (SPP), for which the field theory is known 
\cite{mp,uraconi}. In order to 
see it, we can consider the web diagram before the deformation, namely 
with finite size 2- and 4-cycles, shown in figure \ref{sppz3}b. From this 
we can obtain the toric diagram, shown if figure \ref{toricsppz3}. 
Using techniques in toric geometry, the diagram 
shows manifestly that the geometry corresponds to a $\IZ_3$ quotient of 
the suspended pinch point singularity (SPP), see appendix \ref{quot} for 
more details.

More explicitly, the SPP singularity can be described as the hypersurface 
in $\IC^4$ given by the equation
\beqa
xy-zw^2=0
\label{thespp}
\eeqa
>From the toric data it is possible to read off that our geometry 
corresponds to the quotient by the $\IZ_3$ generated by
\footnote{Although a systematic construction and identification of 
quotients in terms of toric data exists (see appendix \ref{quot}), there is a 
simple 
way to realize 
this is the right symmetry. It is the only $\IZ_3$ symmetry of the SPP 
geometry which has the origin as the only fixed point, in agreement with 
the fact that the additional lattice points in \ref{toricsppz3}b do not 
lie on the sides of the SPP polygon.} 
\beqa
x\to \alpha x \quad , \quad y\to \alpha y\quad ,\quad z\to \alpha z \quad 
, \quad w\to \alpha^2 w
\label{z3action}
\eeqa
with $\alpha=e^{2\pi i/3}$.
Notice that the $\IZ_3$ action leaves invariant the holomorphic 3-form 
$\Omega=\frac{dxdy dz}{zw}$ of the SPP, guaranteeing that the quotient 
is a new CY singularity.

\subsubsection{Geometry of the throat}

The SPP geometry has a nice one-parameter complex deformation to a 
completely smooth space, as described in appendix \ref{appspp} 
\cite{fhu,oz}. This is given by modifying equation (\ref{thespp}) to
\beqa
xy-zw^2=\epsilon w
\label{thedeformedspp}
\eeqa
This is invariant under the $\IZ_3$ action (\ref{z3action}), hence the 
deformation is inherited to the quotient of the SPP.
This corresponds precisely to the geometric separation of the line and the 
$\IC^3/\IZ_3$ sub-web in figure \ref{sppz3}. The geometry contains a 
finite size 3-cycle (which is in fact a Lens space $\IS^3/\IZ_3$), which 
corresponds to the segment joining the two separated sub-webs. The left 
over geometry after the deformation contains a $\IC^3/\IZ_3$ singularity.

On general grounds it is expected that turning on $M$ units of RR flux on 
the finite-size 3-cycles (as well as a suitable NSNS flux on its dual 
(non-compact) 3-cycle) leads to a warped throat. At its 
bottom, the throat is cut off by the finite-size 3-cycles, and it contains 
a $\IC^3/\IZ_3$ singularity. A chiral gauge sector at this bottom is 
obtained by introducing a small set of D3-branes at this singularity
\footnote{It is also possible to introduce anti-D3-branes, as we 
briefly comment in section \ref{further}, but we stick to the 
supersymmetric situation.}. The latter are considered as probes, and do 
not modify the structure of the throat substantially.

The gauge theory on these branes thus leads to a chiral gauge theory
\beqa
 {\rm Vect. mult} \quad & \quad U(N)\times U(N)\times U(N) & \nonumber \\
 {\rm Ch. mult} \quad & \quad 3\, \times \, [\, (N,{\ov N},1) + (1,N,{\ov 
N}) + ({\ov N},1,N)\,] &
\eeqa
where two of the $U(1)$ factors are anomalous and become 
massive by the Green-Schwarz mechanism \cite{sagn,iru} \footnote{
The story for $U(1)$'s in systems of branes at singularities is general. 
All $U(1)$ linear combinations except a diagonal one become massive due to 
$B\wedge F$ couplings. Hence one may work in the quiver theory without 
$U(1)$'s or keep them with the understanding that they eventually 
disappear. We work in these two pictures interchangeably.}. Also, the 
ranks of the three gauge factors are equal due to tadpole/anomaly 
cancellation. Following \cite{aiqu}, in section \ref{d7branes} we 
achieve non-equal ranks by introducing additional D7-branes in the 
configuration. Hence the above theory can be considered a toy model for 
the more realistic SM-like configurations to come.

The explicit metric for the deformed SPP is not known. Hence, it is 
difficult to be more precise about the detailed structure of the throat.
Nevertheless from the geometric viewpoint, the size of the 3-cycle is 
stabilized by fluxes \cite{drs,gkp}. A precise determination of this size 
in terms of the underlying parameters (the fluxes) would require 
determining the dependence of the periods of the holomorphic 3-form with 
the complex structure, around the point in moduli space where the size 
vanishes, as done for the deformed conifold in \cite{gkp}. Even without 
this information, it is reasonable to expect the parametric dependence of 
the size, and the warp factor at the bottom of the throat, to be similar 
to the conifold case. Namely, the ratio of the warp factors at the tip, as 
compared with that at a radial distance $r$ is $\simeq e^{K/Mg_s}$, where 
$N\simeq KM$ is the RR 5-form flux (and 5d cosmological constant) over 
the 5d horizon at the radial distance $r$.

\subsubsection{Holographic field theory dual}
\label{fieldtheorydual}

{\bf The field theory}

The holographic dual description of the above throat is in terms of the 
gauge field theory of D3-branes at our space $Y$, in the singular limit.
In order to construct it, the most efficient way is to exploit its 
realization at a $\IZ_3$ quotient of the SPP quiver gauge theory,
following ideas in \cite{uraconi}.

The gauge group on the world-volume of D3-branes at the SPP singularity is 
$U(N_1)\times U(N_2)\times U(N_3)$, with the chiral multiplet content 
given by
\begin{center}
\begin{tabular}{cccc}
& $U(N_1)$ & $U(N_2)$ & $U(N_3)$ \\
$F$ & $\fund$ & $\antifund$ & $1$ \\
${\widetilde F}$ & $\antifund$ & $\fund$ & $1$ \\
$G$ & $1$ & $\fund$ & $\antifund$  \\
${\widetilde G}$ & $1$ & $\antifund$ & $\fund$  \\
$H$ & $\antifund$ & $1$ & $\fund$  \\
${\widetilde H}$ & $\fund$ & $1$ & $\antifund$  \\
$\Phi$ & Adj. & $1$ & $1$
\label{sppfieldtheory}
\end{tabular}
\end{center}

The ranks are unconstrained by anomaly cancellation, corresponding to the 
possibility of introducing fractional D-branes in the system.

In order to identify the $\IZ_3$ action on the above fields, we note that 
the action (\ref{z3action}) on $x$, $y$, $z$, $w$ corresponds to the 
action
\beqa
& F\to \alpha^2 F \quad ; \quad {\widetilde F}\to {\widetilde F} 
\quad : \quad G\to G \quad ; \quad {\widetilde G}\to \alpha {\widetilde G} 
\nonumber \\
& H\to \alpha^2 H \quad ; \quad {\widetilde H}\to {\widetilde H} \quad ; \quad 
\Phi \to \alpha \Phi
\label{z3actiononfields}
\eeqa
This is obtained as the $\IZ_3$ symmetry of the theory which agrees 
with the action (\ref{z3action}) upon use of the relations (\ref{mapping})
(arising from the realization of the SPP geometry as the moduli space of 
the gauge theory). Alternatively it can be obtained explicitly by using 
the toric techniques in appendix \ref{quot}.

In addition, we need to specify the $\IZ_3$ action on the gauge 
(Chan-Paton) degrees of freedom. This is done by choosing three commuting 
$U(n_i)$ gauge transformations, which without loss of generality can be 
parametrised as the diagonal matrices
\beqa
\gamma_{\theta,1} & = & \diag (\id_{m_0}, \alpha \id_{m_1},\alpha^2 
\id_{m_2})\nonumber \\
\gamma_{\theta,2} & = & \diag (\id_{n_0}, \alpha \id_{n_1},\alpha^2 
\id_{n_2})\nonumber \\
\gamma_{\theta,3} & = & \diag (\id_{p_0}, \alpha \id_{p_1},\alpha^2 
\id_{p_2})
\eeqa

In order to describe the field theory on D3-branes at the quotient space, 
we project the field theory of the SPP onto states 
invariant under the combined (geometric plus Chan-Paton) $\IZ_3$ action.

\begin{figure}[ht]
  \epsfxsize = 10cm
  \centerline{\epsfbox{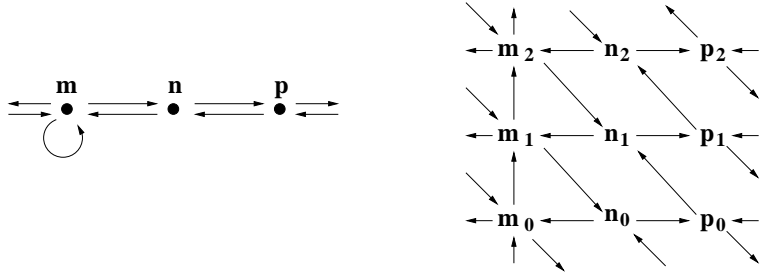}}
  \caption{\small The quiver for the SPP and its orbifold. The SPP quiver 
is understood to be periodically identified along the horizontal 
direction, 
while that of its orbifold should be identified along the horizontal and 
vertical directions.}
  \label{orbifoldspp}
\end{figure}

Regarding the different fields as matrices, the projection conditions on 
the vector multiplets $V_i$, and the chiral multiplets are given by
\beqa
& V_i=\gamma_{\theta,i} V_i \gamma_{\theta,i}^{-1} \quad ; \quad 
 \Phi = \alpha \gamma_{\theta,1} \Phi \gamma_{\theta,1}^{-1} \quad ; \quad
 F= \alpha^2 \gamma_{\theta,1} F\gamma_{\theta,2}^{-1} \quad ; \quad
{\widetilde F}= \gamma_{\theta,2}{\widetilde F} \gamma_{\theta,1}^{-1}
\nonumber \\
& G= \gamma_{\theta,2} G\gamma_{\theta,3}^{-1} \quad ; \quad
{\widetilde G}= \alpha \gamma_{\theta,3}{\widetilde G} \gamma_{\theta,2}^{-1}
\quad ; \quad H= \alpha^2 \gamma_{\theta,3} H\gamma_{\theta,1}^{-1} \quad 
; \quad {\widetilde H}= \gamma_{\theta,1}{\widetilde H} \gamma_{\theta,3}^{-1}
\quad
\eeqa
The quiver gauge theory gauge group is $\prod_i (U(m_i)\times 
U(n_i)\times U(p_i))$, and the chiral multiplet content is given 
by
\beqa
& U(m_i)\times U(n_i)\times U(p_i) \nonumber\\
F_{i,i-1} & (m_i,\ov{n}_{i-1}) \nonumber \\
{\widetilde F}_{i,i} & (n_i,\ov{m}_i) \nonumber \\
G_{i,i} & (n_i,\ov{p}_i) \nonumber \\
{\widetilde G}_{i,i+1} & (p_i,\ov{n}_{i+1}) \nonumber \\
H_{i,i-1} & (p_i,\ov{m}_{i-1}) \nonumber \\
{\widetilde H}_{i,i} & (m_i,\ov{p}_i) \nonumber \\
\Phi_{i,i+1} & (m_i,\ov{m}_{i+1})
\label{finaltheory} 
\eeqa
With this notation, the superpotential reads
\beqa
W & = & {\widetilde F}_{i,i} F_{i,i-1} G_{i-1,i-1} {\widetilde G}_{i-1,i} \, 
-\, {\widetilde G}_{i,i+1} G_{i+1,i+1} H_{i+1,i} {\widetilde H}_{i,i}\, + 
\nonumber \\
& + & {\widetilde H}_{i,i} H_{i,i-1} \Phi_{i-1,i} \, -\, \Phi_{i,i+1} 
F_{i+1,i} {\widetilde F}_{i,i}
\label{finalsupo}
\eeqa
The quiver for the spp and its orbifold are shown in figure 
\ref{orbifoldspp}. 

The realization of our geometry as a quotient of the SPP is 
a useful tool in this example, both in the construction of the quiver, and 
in the subsequent analysis of the field theory. However, we prefer to 
carry out the remaining analysis of the field theory without using this 
information, in order to illustrate that the holographic dual cascade etc 
can be constructed once the quiver gauge theory is known, even for 
examples which are not orbifolds. The relation to similar phenomena in the 
parent theory is described in appendix A.

\medskip

{\bf The duality cascade}

Imposing the conditions of cancellation of anomalies, the ranks of the 
nine groups in the quiver are partially constrained. In the absence of 
further branes in the configuration (e.g. D7-branes like in section 
\ref{d7branes}) the most general rank assignment is
\beqa
m_i=N \quad, \quad n_i=N+M \quad, \quad p_i=N+P
\eeqa
This shows that the configuration can support two kinds of fractional 
branes. In the following we study the behaviour of the field theory for
$P=0$, and show that it leads to a duality cascade. The whole analysis may 
be regarded as a $\IZ_3$ quotient of that in section \ref{sppcascade}.
but we repeat it on its own, since it illustrates the general technique, 
valid for more general examples.

\begin{figure}[ht]
  \epsfxsize = 10cm
  \centerline{\epsfbox{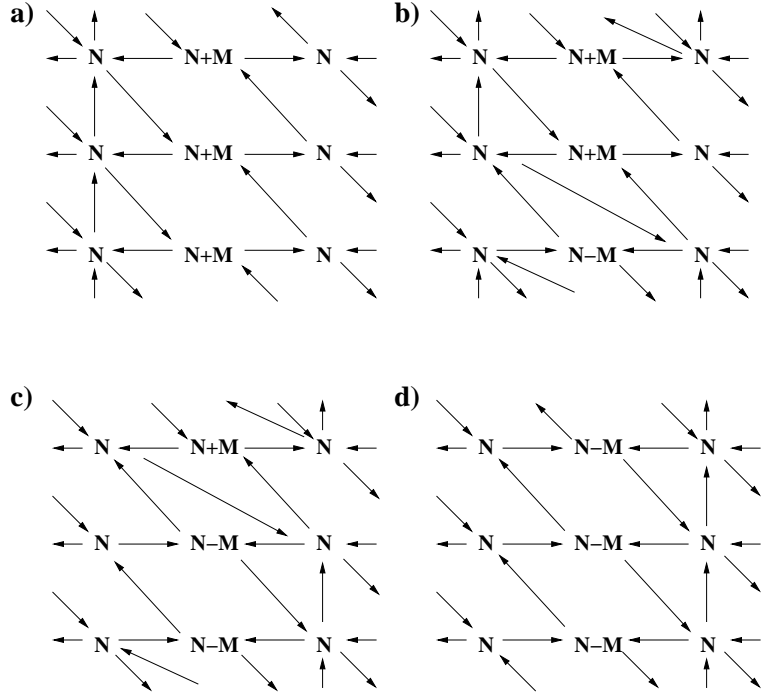}}
  \caption{\small The quivers in the sequential Seiberg dualities. The 
diagrams are to be periodically identified along the horizontal and vertical 
directions. Figure a) shows the initial quiver for the orbifold of the 
SPP. Figure b) shows the quiver after dualising the node $n_0$, 
figure c) shows the quiver after dualising the node $n_1$ and figure d) 
shows the quiver after dualising $n_2$. In each dualization, the dualised 
node ends up with rank $N-M$.}
  \label{dualization}
\end{figure}

We consider the situation where $m_i=N$, $n_i=N+M$, $p_i=N$. We also 
consider the UV couplings of groups in the same row to be equal.
In this 
situation, when the theory flows to the IR we expect the nodes with rank 
$N+M$ to become strongly coupled and we should Seiberg dualise them. This 
may be done sequentially, namely we first dualise $n_0$, then $n_1$ and 
finally $n_2$. As described in \cite{fhhu} (see also \cite{bp}), 
dualization of a node simply amounts to simple operations with the arrows 
of the quiver. Namely, one reverses all arrows with one endpoint in the 
dualised node (replaces quarks by dual quarks), and adds arrows 
corresponding to the composition of ingoing and outgoing arrows 
(introduces the mesons). In addition, one should replace the number of 
colors by its dual ($N_c'=N_f-N_c$) and removes paired arrows (from mass 
terms in the superpotential). The whole process of the three 
Seiberg dualities is shown in figure \ref{dualization}.

The final result is the quiver shown in figure \ref{dualization}d.
It is a quiver of the same kind as the original one, up to a translation 
of the nodes (of two nodes to the left in the horizontal direction), and 
a change of ranks. This facilitates the next dualization steps, which 
follow an analogous pattern.

Following the flow to the infrared, the next nodes in becoming strongly 
coupled are the three nodes of type $m$ (first column), which should be 
simultaneously dualised. Then one should dualise nodes of type $p$, then $n$ 
again, and continue dualising nodes of type $m$, $p$ and $n$ again. After 
this the theory comes back to exactly the original one, with the same 
value of $M$ but with an effective $N$ of  $N-6M$. This completes a 
duality cycle, whose repeated occurrence in the RG flow originates the 
duality cascade.

In section \ref{sppcascade} we compare this flow with that for the SPP 
parent theory with a similar assignment of fractional branes. The 
above flow amounts to simply a $\IZ_3$-projected version of the duality 
cascade of the SPP, studied in \cite{fhu}. This relation guarantees even 
more strongly that the above duality cascade in the orbifold theory takes 
place in the RG flow of the theory, at least for a general 
$\IZ_3$-invariant choice of UV gauge couplings.

\medskip

{\bf The infrared deformation}

The duality cascade proceeds until the effective number of D3-branes is
comparable with the number of fractional branes $M$. In this situation, 
strong infrared dynamics takes over and ends the cascading behaviour. For 
instance, for $N$ a multiple of $M$ the final gauge theory simply 
contains three decoupled $SU(M)$ SYM-like theories, which have a common 
dynamical scale $\Lambda$, at which confinement occurs. In the 
supergravity dual, the warped throat describing 
the cascade is cutoff in the infrared by a complex deformation of the 
geometry, namely by a finite size 3-cycle, to a $\IC^3/\IZ_3$. 

A simple way to recover this deformation from the field theory viewpoint 
is to consider introducing $M$ additional D3-branes probing the infrared 
theory, and to study their moduli space (which corresponds to probing the 
geometry of the supergravity dual). Hence, following \cite{fhu}, we are 
led to studying the quiver gauge theory with $m_i=M$, $n_i=2M$, $p_i=M$. 

In this situation, the nodes $SU(2M)$ have the same number of colors and 
flavours. Hence, ignoring the $SU(M)$ dynamics (whose coupling is smaller, 
hence may be regarded as global symmetries), these gauge factors develop a 
quantum deformation of their moduli space. Specifically, for the 
$i^{th}$ $SU(2M)_i$ factor we introduce the mesons
\beqa
{\cal M}_{i} = 
\left[ \begin{array}{cc} A_{i+1,i} \, & \, B_{i+1,i} \cr
C_{i-1,i}\, & \, D_{i-1,i} \end{array} \right] \, =\,
\left[ \begin{array}{cc} F_{i+1,i}G_{i,i}\, &\, F_{i+1,i}{\widetilde F}_{i,i} \cr
{\widetilde G}_{i-1,i}G_{i,i}\, &\, {\widetilde G}_{i-1,i} {\widetilde F}_{i,i}
\end{array} \right] 
\label{mesons}
\eeqa
and the baryons, schematically ${\cal B}_i$, ${\widetilde{\cal B}}_i$. These 
are the infrared degrees of freedom after confinement of the corresponding 
gauge factor. They are subject to the quantum modified constraint 
\beqa
\det{\cal M}_i - {\cal B}_i{\widetilde {\cal B}}_i=\Lambda^{4M}
\eeqa
where $\Lambda$ is a strong coupling dynamics scale. These constraints can 
be implemented in the superpotential via Lagrange multipliers $X_i$. The 
superpotential thus reads
\beqa
W & = & D_{i-1,i} A_{i,i-1} \, -\, C_{i,i+1} H_{i+1,i} {\widetilde H}_{i,i} 
\, +\, {\widetilde H}_{i,i} H_{i,i-1} \Phi_{i-1,i} \, -\, \Phi_{i,i+1} 
B_{i+1,i}\,+ \nonumber \\
& - & X_i\, (\, \det{\cal M}_i -{\cal B}_i{\widetilde {\cal B}}_i)
\eeqa
The dynamics of the D3-brane probes is manifest along the mesonic branch, 
where the quantum constraint is saturated with vevs for the mesonic operators, 
namely
\beqa
X_i=1 \quad ;\quad {\cal B}_i={\widetilde {\cal B}}_i=0 \quad ; \quad 
\det{\cal M}_i=\Lambda^{4M}
\label{mesonicbranch}
\eeqa
Along this mesonic branch gauge symmetry is broken. The maximally 
unbroken symmetry follows e.g. for $\det{\cal M}_i\propto \id_{4M}$ where 
the breaking pattern is (keeping the rank notation general, for the sake 
of clarity)
\beqa
U(m_0)\times U(p_2)\to U(N_0) \quad ; \quad
U(m_1)\times U(p_0)\to U(N_1) \quad ; \quad
U(m_2)\times U(p_1)\to U(N_2) \quad \quad
\label{breaking}
\eeqa
As discussed in \cite{fhu}, it is permissible (and even convenient for 
future discussion) to include the $U(1)$ centre of mass factors in the 
discussion. 

With respect to the symmetry (\ref{breaking}), the quantum numbers for 
the matter multiplets are
\begin{center}
{\small
\begin{tabular}{cccccc}
$A_{02}$, $D_{20}$: & $(N_0,{\ov N}_0)$ &
$A_{10}$, $D_{01}$: & $(N_1,{\ov N}_1)$ &
$A_{21}$, $D_{12}$: & $(N_2,{\ov N}_2)$ \\
$B_{10}$: & $(N_1,{\ov N}_0)$ &
$B_{21}$: & $(N_2,{\ov N}_1)$ &
$B_{02}$: & $(N_0,{\ov N}_2)$ \\
$C_{20}$, $H_{21}$, $\widetilde{H}_{00}$, $\Phi_{01}$: & $(N_0,{\ov N}_1)$ \quad &
$C_{01}$, $H_{02}$, $\widetilde{H}_{11}$, $\Phi_{12}$: & $(N_1,{\ov N}_2)$ \quad &
$C_{12}$, $H_{10}$, $\widetilde{H}_{22}$, $\Phi_{20}$: & $(N_2,{\ov N}_0)$ 
\end{tabular}
} 
\end{center}
Restricting to the abelian case, and along the mesonic branch, the 
superpotential reads
\beqa
W & = & D_{i-1,i} A_{i,i-1} \, -\, C_{i,i+1} H_{i+1,i} {\widetilde H}_{i,i} 
\, +\, {\widetilde H}_{i,i} H_{i,i-1} \Phi_{i-1,i} \, -\, \Phi_{i,i+1} 
B_{i+1,i}\,+ \nonumber \\
& - & A_{i+1,i} D_{i-1,i} +C_{i-1,i} B_{i+1,i}
\eeqa
Using the equations of motion for the $D$, $A$ and $B$ fields, we obtain
\beqa
& A_{0,2}=A_{1,0}=A_{2,1}\equiv A \nonumber \\
& D_{0,1}=D_{1,2}=D_{2,0}\equiv D \nonumber \\
& \Phi_{1,2}=C_{0,1}\, , \, \Phi_{2,0}=C_{1,2}\, ,\, \Phi_{0,1}=C_{2,0}
\label{freezing}
\eeqa 
Essentially, the fields $B$ become massive with the fields $\Phi-C$, while 
overall combinations of the fields $A$, $B$, $\Phi+C$, $D$ remain light. 
They are subject to the quantum constraint, and describe the motion of the 
probes in a complex deformed geometry. 

The left over geometry is encoded in the left over field theory.
It is useful to relabel the fields as $C_{i-1,i}=X_{i,i+1}$, 
$H_{i,i-1}=Y_{i-2,i-1}$, ${\widetilde H}_{i,i}=Z_{i,i+1}$. The field theory 
has a gauge group $U(N_0)\times U(N_1)\times U(N_2)$, and chiral 
multiplets $X_{i,i+1}$, $Y_{i,i+1}$, $Z_{i,i+1}$ in the representation 
$(N_i,{\ov N}_{i+1})$. Namely
\beqa
3\, \times \, [\, (N_0,{\ov N_1}) + (N_1,{\ov N}_2) + (N_2,{\ov N}_0)\,]
\eeqa
Using (\ref{freezing}) the superpotential reads
\beqa
W & = & -X_{i,i+1} Y_{i+1,i+2} Z_{i+2,i} \, +\, X_{i,i+1} Z_{i+1,i+2} 
Y_{i+2,i}
\eeqa
This is precisely the field theory of D3-branes at the $\IC^3/\IZ_3$ 
singularity. Hence the latter is the left over geometry after the complex 
deformation, namely the structure of the geometry at the end of the 
dual warped throat is a $\IZ_3$ orbifold singularity.

\medskip

Using this information, and following similar discussions of the conifold 
case in \cite{ks} and appendix \ref{appspp} for the SPP, we may consider 
the field theory cascade with $N=kM+P$, with $k\in\IZ$ and $0\leq P\leq M$. 
Again the theory suffers a duality cascade and infrared confinement, 
leaving an $U(P)^3$ theory in the infrared. This can be obtained by 
analysing the next to last step in the cascade, where the ranks are
$m_i=P$, $n_i=M+P$, $p_i=P$. The strong $SU(M+P)$ dynamics confines and 
generates an Affleck-Dine-Seiberg superpotential \cite{ads}. This forces 
the meson 
determinant to acquire a non-zero vev, which triggers a breaking of the 
symmetry and a collapse of the quiver theory exactly as in the above 
analysis. The left over theory indeed describes a $U(P)^3$ gauge theory
associated to the quiver of $P$ D3-branes at the $\IC^3/\IZ_3$ 
singularity. In the dual supergravity description, we have a throat based on 
the quotient of the SPP geometry, which is cutoff in the infrared by a 
3-cycle which corresponds to the complex deformation to the $\IC^3/\IZ_3$ 
singularity, at which we have $P$ explicit D3-brane probes. In heuristic 
terms the dynamics of the $kM$ D3-branes create the complex deformation, 
while the additional $P$ D3-branes remain as probes of the left over 
$\IZ_3$ orbifold singularity.

\medskip

Considering this chiral gauge theory as a momentarily reasonable toy 
model for particle physics, we have provided a purely 4d field theory 
interpretation of the stabilization of the hierarchy, and the appearance 
of the SM fields in the infrared. The Standard Model fields (gauge 
bosons, chiral fields, and Higgsses) are composites of a larger gauge 
theory at higher energies, which is almost conformal in large ranges of 
energies, but eventually confines at a scale (which one should consider 
close to the TeV). The infrared dynamics of this theory is such that (the 
above toy version of) the SM arises as the corresponding composite gauge 
bosons, and mesonic degrees of freedom. Thus the hierarchy is generated 
by a peculiar version of walking technicolor. It is worthwhile to make 
two comments on the peculiarity of this theory: a first unusual feature 
of this model is that the SM gauge bosons are also composites, and arise 
only as infrared excitations. A second one is the cascading structure of 
the UV theory, with different effective theories in different energy 
ranges, each of them being described in terms of degrees of freedom which 
are composite in terms of the theory at immediately higher energies. 
Moreover, there is no energetic regime in which any of these degrees of 
freedom are weakly coupled (notice that there is no UV asymptotic freedom 
in the theory), hence the main signature at high energies is the 
appearance of glueballs of the confining theories at energies immediately 
above the SM. Note that in the supergravity picture these glueballs are 
the Kaluza-Klein excitations of the graviton.

Although more involved (and realistic) models will be described in coming 
sections, the basic physical ideas about the 4d interpretation of the SM 
remain valid in such more general configurations. 

\subsection{Adding D7-branes}
\label{d7branes}

We have succeeded in constructing a throat with a chiral D-brane sector 
at its tip, and with a tractable holographic dual. However, D-brane 
configurations leading to theories with SM-like spectrum require additional 
ingredients, like D7-branes \cite{aiqu,cgqu}. In this section we describe 
the introduction of D7-branes in the warped throat picture, as well as 
the corresponding holographic field theory description.

As discussed  in \cite{kk} and below, the D7-branes may be treated in the 
probe approximation as long as their number is much smaller than the 
fluxes. Also, in order to preserve supersymmetry \footnote{In the 
presence of fluxes, the same condition ensures that the 4-cycle is a 
generalized calibration, and thus that the brane preserves supersymmetry 
\cite{cascurcalibra}.}, the 4-cycle must be holomorphic.

We would like to consider a set of D7-branes, spanning a (non-compact) 
4-cycle passing through the $\IC^3/\IZ_3$ orbifold singularity. These may 
be obtained by considering D7-branes in the SPP geometry, wrapped on a 
4-cycle passing through the $\IZ_3$ fixed point $x=y=z=w=0$, and 
performing the $\IZ_3$ quotient. This 4-cycles should be holomorphic, 
namely, defined by a holomorphic equation in the complex variables $x$, 
$y$, $z$, $w$. 

Again, the difficulty lies in identifying the open string degrees of 
freedom arising from this new ingredient. 
This information may be obtained by starting with D3/D7-brane systems at 
$\IC^2/(\IZ_2\times \IZ_2)$ and carrying out a partial resolution to the 
SPP. Fortunately this kind of exercise has been carried out (in the 
presence of additional orientifold projections) in \cite{pru}, hence we 
may extract enough information for our purposes from this reference. 
In this section we consider the particular example, enough for our 
purposes, of a stack of D7-branes wrapped on the 4-cycle $w=0$ 
\footnote{From the equations of the SPP, this implies $z$ arbitrary and 
$xy=0$. Hence the 4-cycle is reducible in two components, corresponding 
to $x=0$, $y$ arbitrary and $y=0$, $x$ arbitrary. The matter content 
described in section \ref{holsm} corresponds to D7-branes in one of these 
component 4-cycles, while for D7-branes in the 
other 4-cycle leads to a different matter content, which incidentally 
appears in a Seiberg dual version along the cascade, see figure 
\ref{seibergexact}c.}. 

\subsubsection{The Standard Model on the throat}

We consider the throat we had before, namely, 3-form fluxes on the 
deformation of the $\IZ_3$ quotient of the SPP to the $\IZ_3$ singularity 
(equivalently, the $\IZ_3$ quotient of $xy-zw^2=\epsilon w$). We also 
choose the RR flux $M$ much larger than the number of branes to be 
introduced below, so that the latter can be treated as probes.

At the singularity, we would like to consider a system of D3- and D7-brane 
probes leading to the semi-realistic models in \cite{aiqu}, in analogy 
with \cite{cgqu}. This is achieved by introducing six D3-brane probes at 
the $\IZ_3$ singularity, and a number of D7-branes wrapped on $w=0$. In 
order to obtain a realistic spectrum, we make the following choice of 
Chan-Paton matrices 
\beqa
\gamma_{\theta,3} & = & \diag (\id_3,\alpha \id_2,\alpha^2 \id_1) 
\nonumber \\
\gamma_{\theta,7} & = & \diag (\alpha \id_3,\alpha^2 \id_6) 
\label{cpassign}
\eeqa
which is consistent with cancellation of RR tadpoles. 

\begin{figure}[ht]
  \epsfxsize = 4cm
  \centerline{\epsfbox{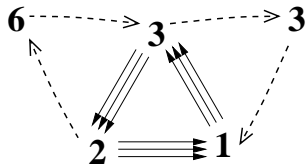}}
  \caption{\small Quiver for the gauge theory at the infrared of the 
cascading throat. It is a 3-family SM-like theory.}
  \label{smquiver}
\end{figure}

Since the model is locally $\IC^3/\IZ_3$, local properties of the 
configuration (like the 33 and 37 spectrum) can be analysed ignoring
the fluxes and the global geometry. The local analysis thus follows 
\cite{aiqu}. As discussed in 
\cite{aiqu}, a single linear combination of the D3-brane $U(1)$´s remains 
light, and plays the role of hypercharge. The group on the D7-branes leads 
to 8d gauge dynamics, which should be considered as a global symmetry from 
the D3-brane viewpoint. Upon embedding the throat in a global 
compactification, the D7-brane gauge dynamics becomes 4d, but the 
unbroken 4d gauge group and matter content depend on the compactification 
boundary conditions and are model-dependent. They will not be discussed 
in our local approach. The spectrum of the model is shown in table 
\ref{tabpssm}, and corresponds to the quiver diagram \ref{smquiver}.

\begin{table}[htb] \footnotesize
\renewcommand{\arraystretch}{1.25}
\begin{center}
\begin{tabular}{|c|c|c|c|c|c|c|}
\hline Matter fields  &  $Q_3$  & $Q_2 $ & $Q_1 $ & $Q_{U(3)}$ &
   $Q_{U(6)}$ & $Y$   \\
\hline\hline {{\bf 33} sector} &  & & & & & \\
%(3 copies) & & & & & & & \\
\hline $3(3,2)$ & 1  & -1 & 0 & 0 & 0 & 1/6  \\
\hline $3(\bar 3,1)$ & -1  & 0  & 1 & 0 & 0 & -2/3 \\
\hline $3(1,2)$ & 0  & 1  & -1 & 0 & 0 & 1/2  \\
\hline\hline {{\bf 37} sector} & & & & & & \\
\hline $(3,1;{\ov 3},1)$ & 1 & 0 & 0 & -1  & 0 & -1/3 \\
\hline $(\bar 3,1;1,6)$ & -1 & 0 & 0 & 0 & 1  & 1/3 \\
\hline $(1,2;1,{\ov 6})$ & 0 & 1 & 0 & 0 & -1 & -1/2 \\
\hline $(1,1;3,1)$ & 0 & 0 & -1 & 1 & 0 & 1 \\
\hline \end{tabular}
\end{center}
\caption{\small Spectrum of $SU(3)\times SU(2)\times U(1)$ model. We
present the quantum numbers  under the $U(1)^5$ groups. The first three
$U(1)$'s come from the D3-brane sector. The next two come from the
D7-brane sectors. The last column shows charges under the D3-brane 
massless $U(1)$ linear combination, which plays the role of hypercharge.
\label{tabpssm} }
\end{table}

The above construction realizes a configuration of D-branes leading to an 
interesting chiral gauge sector localized at the IR end of a warped throat.
The construction is supersymmetric, although one may attempt to do 
non-supersymmetric model building.

\subsubsection{Holography and the Standard Model}
\label{holsm}

We now turn to a description of the holographic dual of the above 
configuration. General results in the gauge/gravity correspondence imply 
that the addition of D7-branes in the gravity side must correspond to the 
addition of flavours in the holographic dual quiver gauge theory. Namely, 
fields in the (anti)fundamental representation of some gauge factors of 
the field theory, and transforming under the global symmetry groups 
associated to the D7-branes. Since the geometry is unchanged, we expect a 
duality cascade of the D3-brane quiver gauge theory, with structure given 
as above (section \ref{fieldtheorydual}). More specifically, the  
$N_f$ D7-branes in the gravity 
side are treated in the probe approximation, namely at zeroth order
$N_f/N$, $N_f/M$. As discussed in \cite{kk}, this corresponds to the 
quenched approximation, in which the effect of the flavours is ignored 
in the gauge dynamics. Hence the gauge dynamics reduces to the 
pure D3-brane case.

Notice that nevertheless, as we describe below, the field theory analysis 
can be carried out even including these effects. This would be 
particularly 
relevant in the infrared in situations where $M$ is not particularly
large.

\medskip

{\bf The field theory}

In the following we discuss the field theory. In order to determine the 37 
fields in the final quiver field theory, we describe them in the SPP 
field theory and carry out the $\IZ_3$ quotient.

The spectrum of 37 strings for the D7-branes on $w=0$ in the SPP can be 
extracted from \cite{pru}. Before the orbifold projection, 
the 3-7 open strings lead to two $\NN=1$ chiral multiplets $Q$, ${\widetilde 
Q}$ in the representations $(N_{D3};{\ov N}_{D7})$ and $({\ov N}_{D3}, 
N_{D7})$, 
where the first entry corresponds to the  D3-brane gauge factor with 
adjoints, and the second to the $U(N_{D7})$ on the D7-branes.
There is also a superpotential
\beqa
W& = & \Phi Q {\widetilde Q}
\label{supo37}
\eeqa

\begin{figure}[ht]
  \epsfxsize = 6cm
  \centerline{\epsfbox{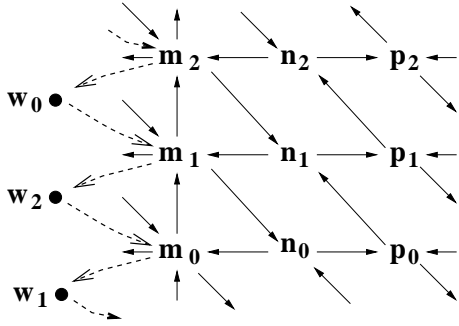}}
  \caption{\small Extended quiver diagram for D3- and D7-branes at the 
quotient of the SPP.}
  \label{extendedquiver}
\end{figure}

It is now easy to quotient the configuration by the $\IZ_3$ symmetry, and 
obtain the effect of the new D7-branes on the quotient field theory. Since 
the D7-branes are mapped to themselves by the $\IZ_3$ action, we 
introduce a general
Chan-Paton action
\beqa
\gamma_{\theta,D7}&=&\diag(\id_{w_0},\alpha \id_{w_1},\alpha^2 \id_{w_2})
\eeqa
The geometric action on $Q$, ${\widetilde Q}$ may be obtained from the 
behaviour of the vertex operators for the corresponding open strings. The 
result is $Q\to \alpha Q$, ${\widetilde Q}\to \alpha {\widetilde Q}$, consistently 
with the $\IZ_3$ invariance of the superpotential (\ref{supo37}). Carrying 
out the $\IZ_3$ projection, the 3-7 spectrum in the quotient of the SPP is
\beqa
{\rm Ch.}\;\; {\rm Mult}\;\; 
\sum_i (m_i; {\ov w}_{i+1}) + ({\ov m}_{i+1}; w_i)
\eeqa
This may be encoded in an extended quiver diagram \cite{dm,uraquiver}, with
nodes associated to the D7-brane groups, and arrows representing the 3-7 
and 7-3 states. This is shown in figure \ref{extendedquiver}.

For notational convenience, we describe the rank assignments in a quiver 
like figure \ref{extendedquiver} as a matrix
\beqa
\pmatrix{w_0 & | & m_2 & n_2 & p_2 \cr
w_2 & | & m_1 & n_1 & p_1 \cr
w_1 & | & m_0 & n_0 & p_0 }
\eeqa
Similarly to the flat space orbifold case, the addition of D7-branes modifies
the RR tadpole cancellation conditions. They can be recovered as the 
conditions of cancellation of non-abelian gauge anomalies 
\cite{abiu}, and will not be further discussed.

\begin{figure}[ht]
  \epsfxsize = 6cm
  \centerline{\epsfbox{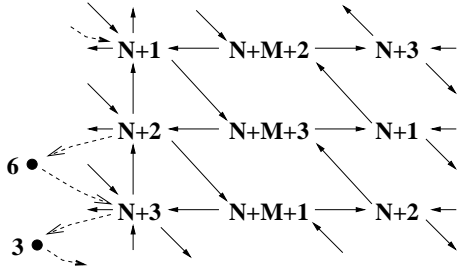}}
  \caption{\small Quiver for the rank assignments leading to the 
duality cascade with flavours.}
  \label{smquiver1}
\end{figure}

In order to describe the holographic dual of the throat with the above 
assignment of D-branes at its tip, we should realize that the D7-branes 
correspond to a global symmetry group, and hence they are unchanged by the 
gauge dynamics. Hence, the choice of ranks should agree with the 
fractional D7-branes on the gravity side, 
namely $w_0=0$, $w_1=3$, $w_2=6$. With this choice, the general 
anomaly free (i.e. RR tadpole free) rank assignment for the D3-brane 
nodes corresponds to the matrix
\beqa
\pmatrix{
0 & | & N_1+1 & N_2+2 & N_3+3 \cr
6& | & N_1+2 & N_2+3 & N_3+1 \cr
3 & | & N_1+3 & N_2+1 & N_3+2 }
\eeqa
Clearly, in order to recover the correct dual gravitational background, we 
are led to the choice $N_1=N_3=N$, $N_2=N+M$, with $N\gg M$. This is shown 
in figure \ref{smquiver1}

Denoting $N_f$ the generic number of flavours, the effect of flavours in 
the field theory is non-trivial, due to the modification of RR tadpole 
conditions, but subleading in $N_f/N$. Hence we expect to recover the 
right structure in the quenched approximation.

\medskip

{\bf The duality cascade}

\begin{figure}[ht]
  \epsfxsize = 14cm
  \centerline{\epsfbox{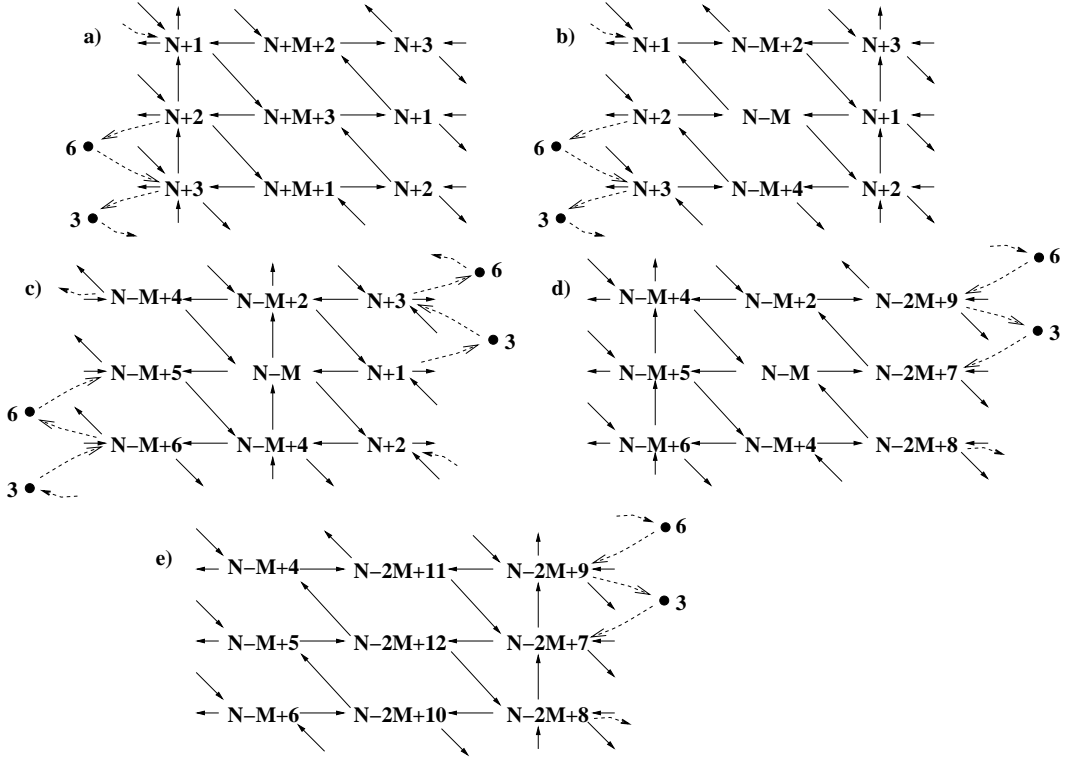}}
  \caption{\small Quivers diagrams for some steps in the duality cascade 
of the theory with flavours. For clarity, in figure c) the D7-brane nodes 
appear replicated.}
  \label{seibergexact}
\end{figure}

The above theory has a duality cascade, which at zeroth order in $N_f/M$ 
reduces to the pure D3-brane cascade. However, there are non-trivial 
effects of the flavours, which we would like to briefly mention. These 
involve changes in the ranks of the dual gauge factors, due to the 
additional flavours, and the appearance of additional meson degrees of 
freedom \footnote{These effects are crucial in keeping the dual theories
anomaly free. In the quenched approximation, some intermediate steps in 
the duality cycle describe anomalous theories, but with anomalies being
subleading in $N_f/N$.}. Several steps in the duality cascade are shown 
in figure 
\ref{seibergexact}. As is manifest by iterating the duality sequence, the 
duality cascade eventually comes back to the original quiver, but with an
assignment of fractional branes differing from the original one in amounts 
of order one. Hence, the cascade is not exactly periodic in the presence 
of flavours. Notice however that these effects are subleading in the 
quenched approximation, hence the geometry in the dual is unchanged with 
respect to the case without flavours. In fact, this discussion has been 
carried in detail for the case of the conifold with flavours 
in \cite{ouyang}, where moreover it was shown that field theory $N_f/N$ 
effects are matched by the 
leading $N_f/N$ backreaction of the D7-brane on the background. Clearly, 
our above throat with the D7-brane regarded as a probe should match the 
cascade in the quenched approximation.

\medskip

{\bf The infrared deformation}

\begin{figure}[ht]
  \epsfxsize = 12cm
  \centerline{\epsfbox{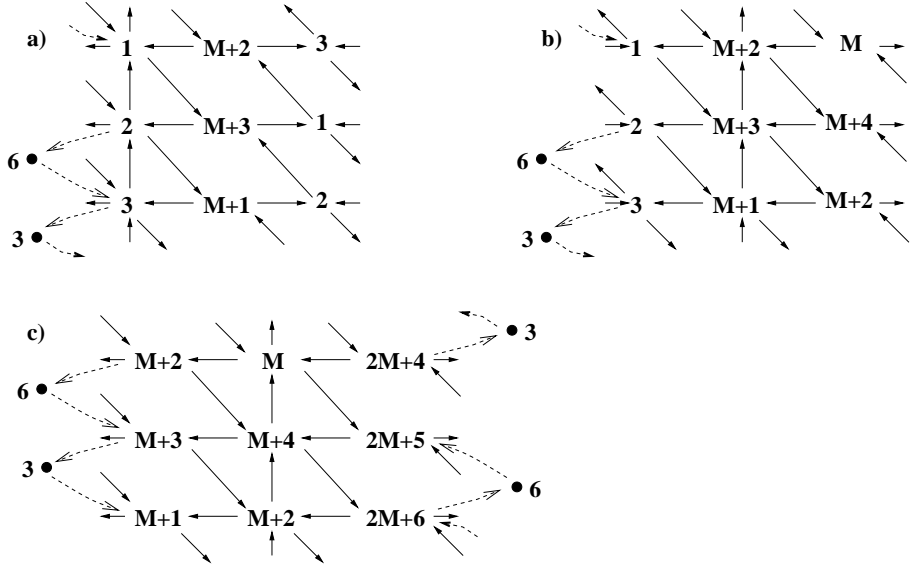}}
  \caption{\small Quivers diagrams for the last steps in the duality 
cascade of the theory with flavours. In going from b) to c) we pushed the diagram one column to the left.}
  \label{smuv}
\end{figure}

The duality cascade ends when the number of D3-branes is exhausted. By 
tuning the choice of UV rank assignments, one may arrange the 
corresponding quiver gauge theory to correspond to the ranks 
\beqa
\pmatrix{
0 & | & 1 & M+2 & 3 \cr
6& | & 2 & M+3 & 1 \cr
3 & | & 3 & M+1 & 2 }
\eeqa
The last steps of the duality cascade can be recovered by taking this IR 
theory as starting point, and dualising up towards the UV. Some of these 
theories are shown in figure \ref{smuv}.

In the supergravity dual we would like to stick to large $M$, so that the 
probe approximation for the D7-branes is valid. From the field theory 
perspective, the system can be analysed even for relatively small $M$. The 
relevant dynamics is as follows. The dynamics is controlled by the 
$SU(M+p_i)$ gauge groups, with other factors only weakly gauged, so they 
are regarded as global symmetries. Thus those gauge factors behave as 
decoupled $SU(N_{c,i})$ gauge theories with $N_{f,i}$ flavours, with 
$N_{f,i}<N_{c,i}$. Concretely the $SU(M+3)$, $SU(M+2)$, $SU(M+1)$ nodes 
have 3, 4, and 5 flavours respectively. Each such node thus 
confines and develops an Affleck-Dine-Seiberg superpotential \cite{ads}.
In terms of the corresponding mesons, which are defined as in 
(\ref{mesons}), the total superpotential reads
\beqa
& W & = D_{i-1,i} A_{i,i-1} \, -\, C_{i,i+1} H_{i+1,i} {\widetilde H}_{i,i} 
\, +\,  {\widetilde H}_{i,i} H_{i,i-1} \Phi_{i-1,i} \, -\, \Phi_{i,i+1} 
B_{i+1,i}\,+ \nonumber \\
& + & N\left(\frac{\Lambda_{SU(N+3)}^{3N+6}}{\det{\cal 
M}_{SU(N+3)}}\right)^{\frac{1}{N}} +
(N-3)\left(\frac{\Lambda_{SU(N+2)}^{3N+1}}{\det{\cal 
M}_{SU(N+2)}}\right)^{\frac{1}{N-3}} +
(N-3)\left(\frac{\Lambda_{SU(N+1)}^{3N-1}}{\det{\cal 
M}_{SU(N+1)}}\right)^{\frac{1}{N-3}} 
\nonumber
\eeqa
As in the case without flavours, the 
non-perturbative superpotential forces the meson determinant to develop a 
vacuum expectation value. The analysis of this theory is somewhat 
analogous to the discussion in section (\ref{fieldtheorydual}), 
after (\ref{mesonicbranch}). The pattern of symmetry breaking is 
(\ref{breaking}) and leads to a $U(3)\times U(2)\times U(1)$ gauge group. 
Computation of the light fields and their superpotential in the effective 
theory gives rise to a gauge theory associated with the quiver in figure
\ref{smquiver}. This is in complete agreement with the supergravity 
picture with the D3/D7-brane probe system.

The above construction realizes an almost conformal cascading gauge theory
which, at an infrared scale, develops confinement leaving a set of 
light degrees of freedom with a structure strikingly close to the Standard 
Model. This is an explicit realization of Strassler's ideas, described 
in \cite{ks,strassler}, on the realization of the SM as the infrared of a 
duality cascade.

Hence the model should be regarded as a (supersymmetric) walking 
technicolor model. Moreover this theory is the holographic 
dual of the supergravity throat described above.

\subsection{Further possibilities}
\label{further}

Our aim has been developing general rules and techniques to construct 
warped throats containing singularities after complex deformation, and 
providing the relevant gauge theory dynamics matching the dual geometry.
Clearly many other models, beyond those considered here, are possible.
In this section we sketch some possibilities.

A simple modification is to consider other local configurations of 
D3-branes leading to semi-realistic models. In particular one may easily 
construct 3-family left-right symmetric models by simply making a 
different choice of Chan-Paton actions on the D3- and D7-branes of the 
model, following \cite{aiqu}. The discussion of the throat and its 
holographic dual cascade is completely analogous to the case we have 
studied. In this construction, the infrared gauge group is $SU(3)\times 
SU(2)_L\times SU(2)_R$, and all quarks and Higgsses are composites, while 
leptons are fundamental (providing a tantalising suggestion of the
quark-lepton Yukawa hierarchy). We leave a more detailed discussion as an 
exercise for the interested reader.

As an alternative kind of embedding, one may embed the SM on sets of 
D7-brane probes instead of 
D3-branes, but leading otherwise to similar spectra (e.g. as given in 
table \ref{tabpssm}, with 33 states in the table arising from 77 states). 
Although we do not describe this in detail, we would like to point that 
in the holographic description, 77 states like gauge bosons and the up 
quarks and up type Higgsses are bulk fields and thus fundamental, while 37 
states like leptons and the down quarks and down type Higgsses sit at the 
bottom of the 
throat and are in fact composite. Thus this setup still may allow an 
explanation of the hierarchy, since e.g. the $\mu$ term is naturally 
suppressed. Moreover, bulk gauge bosons have been argued as a useful setup 
to address gauge coupling unification \cite{unification}, and to suppress 
radiative corrections (by introducing a custodial $SU(2)$  
\cite{custodial} e.g. 
by constructing left-right symmetric models, eventually broken by boundary 
conditions in the ultraviolet, namely the bulk of the Calabi-Yau 
compactification). 

Another interesting possibility would be to embed the SM in a sector of 
anti-D-branes, which leads to non-supersymmetric models. In the 
supergravity picture this is straightforward, and the relevant techniques 
can be adapted from the local analysis in \cite{aiqu}. The holographic 
field theory description should involve a generalization of the 
non-supersymmetric baryonic branch studied in \cite{kpv} for the conifold 
theory. One advantage of models with anti-D3-branes, as emphasised in 
\cite{cgqu} is that they naturally fall to the bottom of warped throats.

We leave these and other extensions for future research.

\section{Multi-warp throats}
\label{multiwarp}

In the previous section we have considered geometries which, after complex 
deformation, leave a non-trivial geometry which nevertheless does not 
admit further deformation. An important alternative class of throats is 
obtained if instead the left over singularity admits further deformations. 
This implies that in the presence of suitable fractional branes, the left 
over singularity develops a further throat, possibly ending in a further 
complex deformation. Such examples were described in \cite{fhu}, and shown 
to correspond to gauge theories with several scales of partial confinement.
In this section we describe general techniques to construct geometrical 
setups of this kind, and briefly sketch their holographic field theory 
description.

\subsection{General construction}

We are interested in warped throats with several radial regions of 
different warp factor.
The complete geometry of such constructions are based on a warped 
version of a local Calabi-Yau geometry, which contains several 3-cycles of 
widely different size. Namely, the sizes of such 3-cycles are stabilized 
at hierarchically different values due to different 3-form fields. We will 
be more precise about this later on for a particular class of examples.

\begin{figure}
\begin{center}
\centering
\epsfysize=5cm
\leavevmode
\epsfbox{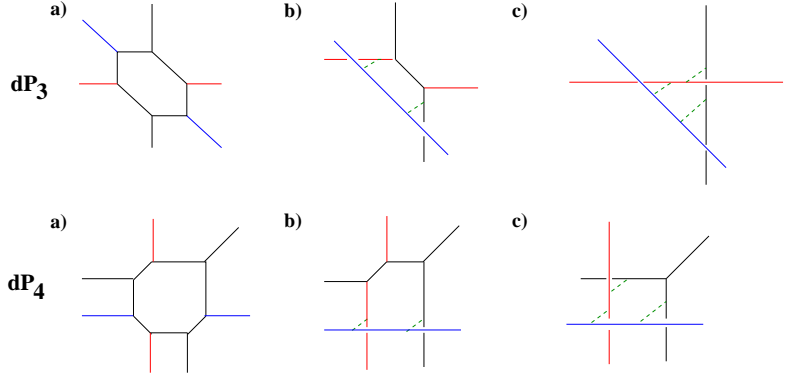}
\end{center}    
\caption[]{\small Two-parameter complex deformations for the complex cone 
over $dP_3$ and the complex cone over (a toric realization of) $dP_4$.}
\label{severaldeform}
\end{figure}

\begin{figure}
\begin{center}
\centering
\epsfysize=3.5cm
\leavevmode
\epsfbox{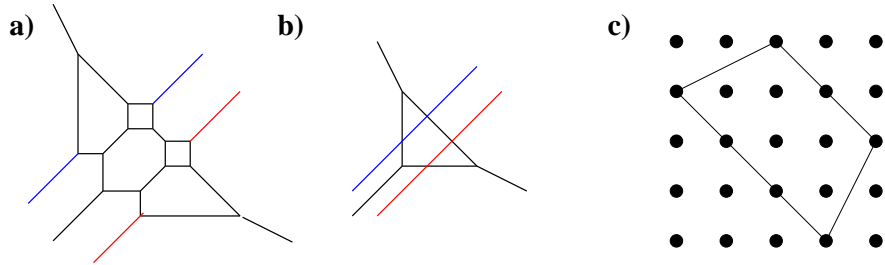}
\end{center}    
\caption[]{\small Web and toric diagrams for a geometry that, supplemented 
with fluxes, describes a two-warp throat ending on a $\IC^3/\IZ_3$ 
singularity. Figure a) and b) describe the generic web diagram, and the 
diagram at the origin of the deformation branch, respectively. Figure 
c) shows the toric diagram, which indicates that the geometry is a 
$\IZ_3$ quotient of the space $xy=z^2w^3$.}
\label{doublethroatz3}
\end{figure}

Hence the starting point we need is a singularity which admits 
several complex deformations. This is easily achieved using the techniques 
in section \ref{deformed}. Namely, we consider geometries associated to 
web diagrams which contain several sub-webs in equilibrium which can be 
removed from the picture. Reference \cite{fhu} studied several examples, shown 
in figure \ref{severaldeform}, with two scales of deformation. It is 
straightforward to construct other geometries where there are more than 
two scales of deformation. Also, one can construct examples where there 
are several scales of deformation, leaving a left over geometry which 
contains a singularity admitting no further deformation. One such example 
is shown in figure \ref{doublethroatz3}. These geometries can therefore be 
used to 
construct throats with several regions of different warp factors, and with 
an infrared end given by a smooth or singular geometry. In the latter 
case, the introduction of D-brane probes leads to chiral gauge theories 
at the bottom of a multi-warp throat, generalising the construction in 
previous section. Figure \ref{doublethroatz3} provides a concrete example 
of one such geometry, which can be used to construct a two-warp throat 
ending on a $\IZ_3$ orbifold singularity, at which one can localise the SM 
chiral 
gauge sector \footnote{Moreover, the geometry corresponds to an orbifold 
of the geometry $xy=z^2w^3$, for which the dual quiver theory can be 
determined using tools in \cite{uraconi}. Hence, the field theory dual, 
the cascade flow and the infrared deformations for the multi-warp 
throat SM may be studied by a combination of the techniques in the 
previous and the coming sections. We leave this exercise for the 
interested reader.}.

Again, as soon as geometries become more involved it is non-trivial to 
construct the corresponding quiver gauge theories, which provide the 
holographic duals of these throats. Nevertheless, some examples of this 
kind have been achieved, and the relevant gauge theory dynamics is 
described in the next section.

\subsection{Holographic description}

In this section we review the simplest such example, studied in 
\cite{fhu}, which illustrates the general gauge theory dynamics 
underlying general multi-warp throats. The geometry is the complex cone 
over $dP_3$, which has a two-parameter set of complex deformations, shown 
in figure \ref{severaldeform}, 
in terms of the web diagram for the geometry (there is an alternative 
one-parameter branch of complex deformations \cite{fhu} which will not 
interest us here).

\begin{figure}
\begin{center}
\centering
\epsfysize=3.5cm
\leavevmode
\epsfbox{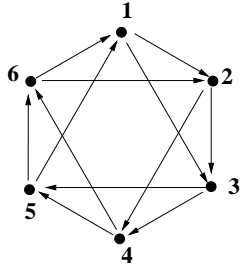}
\end{center}    
\caption[]{\small The quiver for D3-branes on the complex cone over
dP$_3$.}
\label{quiverdp3}
\end{figure}

\begin{figure}
\begin{center}
\centering
\epsfysize=4.3cm
\leavevmode
\epsfbox{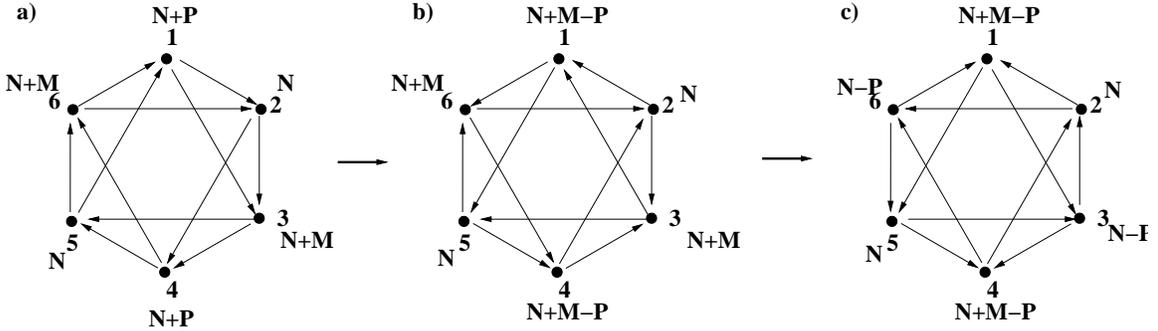}
\end{center}
\caption[]{\small Two dualisations in the first RG cascade in dP$_3$.}
\label{cascade1}
\end{figure}

The quiver theory for D3-branes at this singularity was described in 
\cite{delpezzo}, and corresponds to the diagram in figure 
\ref{cascade1}a. The superpotential is
\beqa
W & = & X_{12} X_{23} X_{34} X_{45} X_{56} X_{61} \, +\,  X_{13} X_{35} 
X_{51} + X_{24} X_{46} 
X_{62} \, - \nonumber  \\
&-& X_{23} X_{35} X_{56} X_{62}\, -\, X_{13} X_{34} X_{46} X_{61}
\, -\,  X_{12} X_{24} X_{45} X_{51}
\eeqa
in self-explanatory notation.

A basis of fractional branes is given by the rank vectors
$(1,0,0,1,0,0)$, $(0,0,1,0,0,1)$  and $(1,0,1,0,1,0)$. The duality cascade 
associated to the two-warp throat, namely to the two-parameter complex 
deformation, corresponds to a choice of ranks
\beq
\vec{N}=N (1,1,1,1,1,1)+P(1,0,0,1,0,0)+M(0,0,1,0,0,1)
\eeq
where we consider $N \, \gg \, P \, \gg \, M$. Given 
the $\IZ_2$ symmetry of the quiver, a symmetric choice of gauge 
couplings in the ultraviolet is preserved under the RG flow, and the 
analysis of the cascade is simplified. The duality cascade proceeds as 
follows. The nodes with largest beta function are 1 and 4, so we dualise 
them simultaneously. The results are shown in Figure \ref{cascade1}a,b. 
Next the most strongly coupled nodes are 3, 6, so we dualise them 
simultaneously, as shown in Figure \ref{cascade1}b,c.

The quiver in Figure \ref{cascade1}c can be reordered into a 
quiver like the initial one, with similar fractional branes, but with the 
effective $N$ reduced by an amount $P$. We can then continue dualising 
nodes 2, 5, then 1, 4, then 3, 6 etc, following the above pattern and 
generating a cascade which preserves the fractional branes but reduces the 
effective $N$.

The cascade proceeds until the effective $N$ is not large compared with 
$P$. For simplicity, consider that the starting $N$ is $N=(k+2)P-M$. Then
after a suitable number of cascade steps, the ranks in the quiver are 
$(2P-M, P-M, P, 2P-M, P-M, P)$, for nodes 123456. In the next duality step 
the theory experiences partial confinement, and the quiver diagram of the 
theory collapses to a simpler one. This is the holographic dual of the 
first complex deformation ending the warped throat in the supergravity 
description.

In order to identify the remaining quiver theory and the corresponding 
geometry, we may consider the mesonic branch of this next-to-last step in 
the duality cascade, which describes the dynamics of D3-branes probes of
the infrared theory. At this stage the $SU(2P-M)$ factors
have $2P-M$ flavours and develop a quantum deformation of their moduli 
space. Following the familiar procedure we introduce the mesons
\beqa
{\cal M}=\left[\begin{array}{cc} M_{63} & M_{62} \\ M_{53} & M_{52} 
\end{array}\right] = 
\left[\begin{array}{cc} X_{61} X_{13} & X_{61} X_{12} \\ X_{51} X_{13} & 
X_{51} X_{12}
\end{array}\right]
\quad ; \quad
{\cal N}=
\left[\begin{array}{cc} N_{36} & N_{35} \cr N_{26} & N_{25} 
\end{array}\right]=
\left[\begin{array}{cc} X_{34} X_{46} & X_{34} X_{45} \\ X_{24} X_{46} & 
X_{24} X_{45}
\end{array}\right]
\nonumber
\eeqa
and the baryons ${\mathcal B}$, ${\widetilde {\mathcal B}}$, 
${\mathcal A}$, ${\widetilde {\mathcal A}}$. As usual we impose the quantum 
constraints via Lagrange multipliers $X_1$, $X_2$, and saturate them 
along the mesonic branch. The meson vevs break the symmetry to 
$SU(P)\times SU(P+M)$ where the first factor arises from nodes 2 and 5, 
and the second from 3 and 6. Restrict to the Abelian case, the 
superpotential reads
\beqa
W & = & M_{62} X_{23} N_{35} X_{56} -  X_{23} X_{35} X_{56} X_{62} - 
M_{63} N_{36}
- M_{52} N_{25}+ \nonumber \\
& + & M_{53} X_{35} + N_{26} X_{62}+ M_{63}M_{52}-M_{53}M_{62} 
+N_{36}N_{25}-N_{26}N_{35}
\label{W_dP3_mesonic}
\eeqa
Using the equations of motion for e.g. $M_{53}$, $N_{26}$, $M_{63}, 
N_{25}$, we are left with
\beqa
W & = & X_{23} N_{35} X_{56} M_{62} - X_{23} M_{62} X_{56} N_{35}
\eeqa
Going back to the non-Abelian case, this is the gauge theory of D3-branes 
at a conifold singularity, showing that the left over geometry after the
complex deformation is a conifold (as expected from the web diagram in 
the top part of figure \ref{severaldeform}. Since the conifold 
theory also contains fractional branes, 
the latter trigger a new duality cascade in which $P$ is reduced 
in $2M$ units in each step \cite{ks}. This leads to a warped throat 
starting at the IR of the previous one, as in figure \ref{twopos}b.

\subsection{Description a la RS}

The general picture is thus a gauge theory with several stages of partial 
confinement, hence corresponding to throats with several warp regions. The 
horizon topology changes in crossing from one region to another, since it 
involves losing or gaining the finite-size 3-cycles in the transitions.
A more detailed supergravity description in the bulk of each warp region 
is possible by simply taking the single-warp version of throats based on 
geometries with the corresponding 5d horizon \cite{ks,chaotic,ejaz}. 
Patching these together would lead to a description of the full throat, 
in the approximation of ignoring the finite extent of the transition 
regions. Unfortunately, even the set of known single warp solutions is 
relatively small.

In this section we provide an even more simplified description of the 
multi-warp throats, which may nevertheless be useful in applications. 
One ignores the internal 5d space and approximates each warp region by a 
slice of an $AdS_5$ of suitable 
radius, and glues them together at distances determined by the flux data of 
the real throat. This is in the spirit of approximating the KS throat by 
an $AdS_5$ geometry with an IR cutoff. 

The description also reproduces RS constructions with several positive 
tension branes (corresponding to our transition regions), hence we may 
simply translate the results from the construction in e.g. \cite{severalRS} 
(modulo trivial notation changes). We consider several `branes' located at 
positions $y_L<y_{L-1}<\ldots <y_1<y_0$, $i=0,\ldots, L$ in the fifth 
coordinate $y$ (with larger $y$ corresponding to further in the IR), 
which extends up to $y=y_0$. At each `brane' we associate 
$M_i$, $K_i$ units of RR and NSNS 3-form fluxes, so that the effective 
D3-brane number for $y_{i+1}<y<y_i$ is $N_i=M_iK_i$. Here and 
henceforth repeated indices are not summed. The metric is given 
by
\beqa
ds^2 & = & e^{-2\sigma(y)}\, ds_{4d}^2 \, +\, dy^2
\eeqa
where $\sigma(y)$ is a continuous piecewise linear function of $y$, with 
slope $1/R_i$ in the interval $y_{i+1}<y<y_i$, with
\beqa
R_i^4\, =\, a_i g_s N_i \alpha'^2
\eeqa
where $a_i$ is a numerical coefficient depending on the 5d horizon in the 
real throat. Finally, in the real throat there is a linear running of the 
effective 5-form flux with $y$
\beqa
{\cal K}(y) \, =\, N_i\, +\, b_i g_s M_i^2 (-y+y_{i+1})/R_i
\eeqa
where $b_i$ is a numerical coefficient depending on the properties of 
3-cycles in the 5d horizon.
Hence, the positions $y_i$ are fixed by the flux data, by the 
condition that each warp region lasts until one runs out of 5-form flux. 
Hence, the boundaries of each warp region are related by
\beqa
e^{-2y_i/R_i}\, =\, e^{-2y_{i+1}/R_i} \, e^{-\frac{K_i}{b_ig_s M_i}}
\eeqa
Clearly the warp factor at the IR of the full throat is related to the UV 
one by a product of exponentials, as expected from the field theory 
analysis.

\subsection{Applications}

The model building applications of multi-warp throats are many, given 
their flexibility in generating several hierarchies. In this section we 
would like to simply point out some possibilities, whose details we leave 
for future research.

A possible application is the construction of brane inflation models 
\cite{inflation,kklmmt} where inflation takes place in the same throat in 
which the SM is located. This is not possible with traditional throats, 
since the warp factor required for inflation and for solving the 
Planck/electroweak hierarchy  are widely different. This can be elegantly 
solved by the use of two-warp throats. Notice that, for models where the 
SM configuration requires D3-branes (as in our case), this scenario 
does not lead stable strings after reheating \cite{strings}.

Another interesting possibility is the construction of inflationary models 
with several stages of inflation. This would manifest in the appearance of 
relatively abrupt changes in the slow-roll parameters, and therefore in 
the observables depending on them. This is nevertheless irrelevant unless 
the transition regions are crossed in the last 60 efoldings (or whatever 
number corresponding to the `observable' stage of inflation).

\medskip

Besides these cosmological applications, there may well be possibilities 
to exploit the multiple hierarchies in a multi-warp throat. Namely, 
using similar ideas it may be possible to construct throats with 
additional  sectors localized in the transition regions. These could be 
exploited to generate new physics at interesting particle physics scales, 
like the intermediate scale or the GUT scale. We leave these interesting 
issues for further work.

\section{Conclusions}
\label{conclusions}

In this paper we have developed techniques to construct warped throats 
with rich geometrical properties, as well as their holographic dual 
descriptions in terms of cascading RG flows ending on partial confinement.

We have proposed two interesting ideas, exploiting these richer 
geometries: on the one hand throats ending on singularities, with D-branes localized on 
them, and leading to chiral gauge sector at the IR end of the throat; on the other, multi-warp throats as in \cite{fhu}.

We have provided a quite explicit description of these systems. In 
particular we have described in detail the holographic dual of a throat 
ending on a system of D3/D7-branes at a $\IC^3/\IZ_3$ singularity, 
realizing a 3-family SM like gauge sector. Independently of the string 
theory setup, the field theory provides the 
construction of an almost conformal UV gauge theory which partially 
confines at low energies, and leaves the SM as its light degrees of 
freedom. This is an explicit realization of the SM as a composite model, 
in the IR of a duality cascade (along ideas in \cite{strassler}).

These results are remarkable. Nevertheless, many open questions 
remain. It would be desirable to find mechanisms to break supersymmetry 
while retaining control of the throat geometry. It would also be 
interesting to adapt these ideas to model building with more flexible 
brane configurations, like intersecting or magnetised D-branes. In this 
respect the local models in \cite{uralocal} could be a good starting 
point. Finally, 
it would be extremely interesting to sharpen the understanding of throats 
that we have, namely by finding additional explicit metrics for horizon 
manifolds, or the details of the backreaction of D7-branes in models 
containing them.

We expect that, based on this work and on a more profound understanding of 
the gauge dynamics underlying warped throats, these and other open 
questions can be addressed, leading to improved model building potential 
for these constructions.

\bigskip

\centerline{\bf Acknowledgements}

We thank A. Delgado, L. E. Ib\'a\~nez, F. Quevedo and R. Rattazzi for 
useful discussions. A. M. U. thanks Ami Hanany and 
Sebastian Franco for collaboration on related subjects, and M. Gonz\'alez 
for kind encouragement and support. J. G. C. and F. S. thank the CERN 
PH-TH Division for hospitality during completion of this work and the spanish Ministerio de Educaci\'on y Ciencia for financial support through F.P.U. grants.
J. G. C. thanks M. P\'erez for her patience and affection.
This work is partially supported by the CICYT, Spain, under project 
FPA2003-02877, and by the networks 
MRTN-CT-2004-005104 `Constituents, Fundamental Forces and Symmetries of 
the Universe', and MRTN-CT-2004-503369 `Quest for Unification'.

\bigskip

\appendix

\section{The SPP and its cascade}
\label{appspp}

\subsection{The SPP field theory and moduli space}
\label{sppmoduli}

\begin{figure}[ht]
  \epsfxsize = 3.5cm
  \centerline{\epsfbox{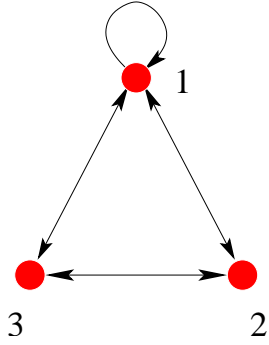}}
  \caption{\small Quiver diagram for SPP.}
  \label{quiver_SPP}
\end{figure}

The gauge theory on D3-branes at a suspended pinch point (SPP) singularity 
\cite{mp,uraconi} has gauge group $U(N_1)\times U(N_2)\times U(N_3)$ and 
matter content 
\begin{center}
\begin{tabular}{cccc}
 & $U(N_1)$ & $U(N_2)$ & $U(N_3)$ \\
$F$ & $\fund$ & $\antifund$ & $1$ \\
${\widetilde F}$ & $\antifund$ & $\fund$ & $1$ \\
$G$ & $1$ & $\fund$ & $\antifund$  \\
${\widetilde G}$ & $1$ & $\antifund$ & $\fund$  \\
$H$ & $\antifund$ & $1$ & $\fund$  \\
${\widetilde H}$ & $\fund$ & $1$ & $\antifund$  \\
$\Phi$ & Adj. & $1$ & $1$
\label{sppfieldth2}
\end{tabular}
\end{center}
The superpotential is
\beqa
W & = & \tr\, \left(
{\widetilde F}FG{\widetilde G} \, - \, {\widetilde G}G H {\widetilde H}
\, +\, {\widetilde H} H\Phi\, -\, \Phi F{\widetilde F} \right)
\eeqa
where bi-fundamental fields are regarded as matrices. The quiver diagram 
is shown in Figure \ref{quiver_SPP}.

When $N_1=N_2=N_3\equiv N$, the generic mesonic branch describes the 
motion of $N$ D3-branes on the SPP geometry. Thus, for $N=1$ the moduli 
space is given by the SPP geometry itself. This can be shown using the 
techniques in \cite{dgm}, as done explicitly in \cite{mp,pru}.

The possible vevs for the above fields are constrained by the F-term 
equations from the superpotential. These can be satisfied automatically if 
we express the above fields in terms of a set of new variables, $p_i$, 
with $i=1,\ldots, 6$, as follows
\beqa
F=p_1p_5\quad , \quad {\widetilde F}=p_2p_3 \quad , \quad G=p_4 \quad , \quad 
{\widetilde G}=p_6 \nonumber \\
H=p_3p_5 \quad , \quad {\widetilde H}= p_1p_2 \quad , \quad \Phi=p_4p_6
\label{intermsofps}
\eeqa
Thus, relations like
\beqa
\frac{\partial W}{\partial \widetilde H}={\widetilde G}GH - H \Phi\, =\, 0, 
{\rm etc}
\eeqa
are automatically satisfied when using (\ref{intermsofps}).

The moduli space is described by gauge invariant operators, modulo F-term 
relations. Namely, combinations of the fields modulo relations obtained 
when expressed in terms of (\ref{intermsofps}).
The corresponding monomials can be chosen e.g. as
\beqa
x & = & {\widetilde H}{\widetilde G}{\widetilde F}=p_1p_2^2 p_3 p_6 \nonumber \\
y & = & FGH=p_1p_3p_4p_5^2 \nonumber \\
z&=&\Phi=p_4p_6 \nonumber \\
w&=&F{\widetilde F}=p_1p_2p_3p_5
\label{mapping}
\eeqa
where of course the F-term relation also allow to say e.g. $z=G{\widetilde 
G}$, etc.

The monomial satisfy the relation
\beqa
xy=zw^2
\eeqa
hence parametrise a moduli space which corresponds to the SPP singularity.

\medskip

In the more technical but powerful language of toric geometry, the moduli 
space is constructed as the symplectic quotient associated to the toric 
data \cite{mp,pru} 
%\begin{center}
\begin{displaymath}
\begin{tabular}{ccccccc}
 & $p_1$ & $p_2$ & $p_3$ & $p_4$ & $p_5$ & $p_6$ \\
$Q_F$ & $1$ & $-1$ & $1$ & $0$ & $-1$ & $0$ \\
$Q_2$ & $-1$ & $1$ & $0$ & $1$ & $0$ & $-1$ \\
$Q_3$ & $-1$ & $0$ & $0$ & $-1$ & $1$ & $1$ 
\label{toriccharges}
\end{tabular}
\end{displaymath}
%\end{center}
Namely the moduli space is parametrised by vevs for the $p_i$'s modulo the 
gauge equivalence by the $U(1)^3$ symmetry with the above 
charge assignment. The $Q_F$ charges are constructed so that the fields of 
the SPP theory are neutral; hence $Q_F$-gauge invariant vevs for the 
$p_i$'s can be associated to vevs for the SPP fields, satisfying the 
F-term constraints. The $Q_2$, $Q_3$ charges impose the D-term constraints 
of the SPP theory at the level of the $p_i$'s. Hence the $U(1)^3$-gauge 
invariant vevs for $p_i$'s correspond to F- and D-flat vevs for the SPP 
theory.

\begin{figure}[ht]
  \epsfxsize = 4cm
  \centerline{\epsfbox{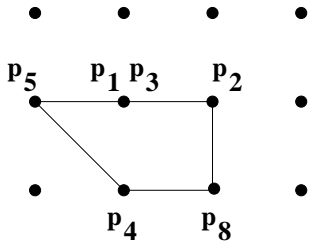}}
  \caption{\small Toric diagram for the SPP geometry.}
  \label{toricspp}
\end{figure}

The toric diagram for the resulting geometry is easily found from 
the toric data above. Namely, the points in the toric diagram have 
coordinates given by the columns in the cokernel matrix
\beqa
T & = & \pmatrix{
0 & 1 & 0 & 0 & -1 & 1 \cr
1 & 1 & 1 & 0 & 1 & 0 \cr
1 & 1 & 1 & 1 & 1 & 1
}
\eeqa
All points lie on a plane, as should be for a Calabi-Yau geometry. The 
points on the plane are shown in figure \ref{toricspp}a

\subsection{The SPP duality cascade}
\label{sppcascade}

The existence of a duality cascade for the SPP field theory in the 
presence of fractional branes has been established and discussed in 
\cite{fhu}.

\begin{figure}[ht]
  \epsfxsize = 11cm
  \centerline{\epsfbox{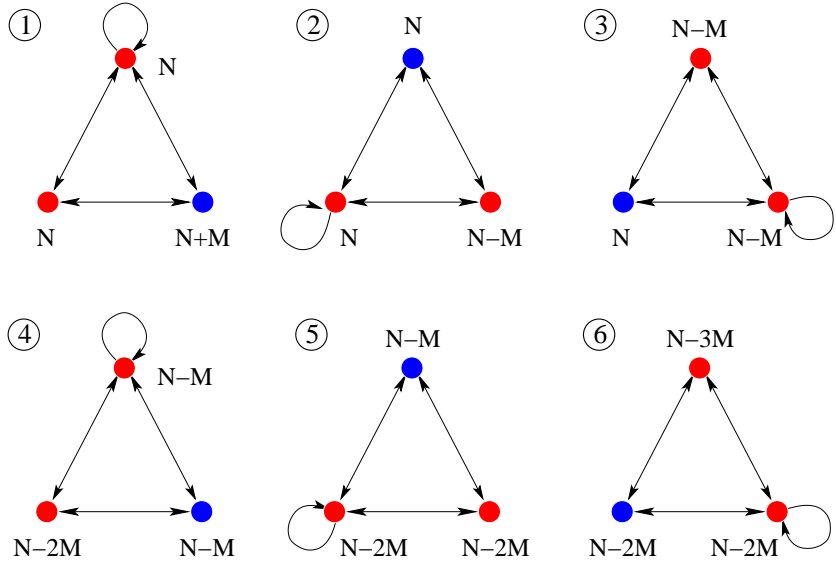}}
  \caption{\small Sequence of quivers in one period of the SPP cascade. We 
have indicated in blue the dualised node at each step.}
  \label{quivers_cascade_SPP}
\end{figure}

The ranks of the gauge factors are arbitrary, hence there are two 
independent fractional branes, which can be taken to be $(0,1,0)$ and 
$(0,0,1)$. Let us consider the starting point given by the ranks  
\beq
\vec{N}=N(1,1,1)+M(0,1,0)
\eeq
By following the pattern of dualising the most strongly coupled node at 
each step, we are led to a cascade that repeats the
following sequence of dualisations $(2,1,3,2,1,3)$. The quiver theories at 
each step of this sequence are shown in figure \ref{quivers_cascade_SPP}. We have indicated in blue the node that gets 
dualised at each step. After six dualisations, the quiver 
comes back to itself, with $N \rightarrow N-3M$ and $M$ constant. This 
pattern of dualisations is consistent with a RG flow, as discussed in 
\cite{fhu}.

\subsection{The infrared deformation}
\label{sppirdef}

\begin{figure}[ht]
  \epsfxsize = 8.5cm
  \centerline{\epsfbox{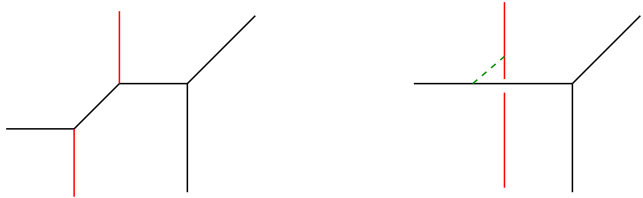}}
  \caption{\small Web diagram for the SPP and its deformation to a smooth 
geometry.}
  \label{web_SPP}
\end{figure}

As usual, the cascade proceeds until the effective number of D3-branes is 
comparable with $M$. After this we expect the gauge theory strong dynamics 
to take over and induce a geometric transition. Indeed, the SPP 
singularity admits a complex deformation. This is manifest using the web 
diagram for the SPP geometry, figure \ref{web_SPP}a, which contains a 
sub-web in equilibrium (a line) which may be removed from the picture, as 
shown in figure \ref{web_SPP}b. In the following we describe how this 
arises in the field theory, following \cite{fhu}.

In order to study the infrared end of the cascade, we study the gauge 
theory describing $M$ D3-branes probing it. This corresponds
to the quiver theory with rank vector
\beq
\vec{N}=M(1,1,1)+M(0,1,0)
\eeq
In this case, we only need to consider mesons for node 2, given by
\beq
{\mathcal{M}} = \left[\begin{array}{cc} A & B \\
C & D\end{array} \right]
            = \left[\begin{array}{ccc} F G & \ & F{\widetilde F} \\ 
{\widetilde G} G & \ & {\widetilde G}{\widetilde F} \end{array} \right]
\label{mesonsspp}
\eeq
and the baryons ${\cal B}$, ${\widetilde {\cal B}}$. The quantum constraint in 
the superpotential reads
\beqa
\det {\mathcal M}-{\cal B}{\widetilde {\cal B}}=\Lambda^{4M}
\eeqa
To derive the dynamics of the $M$ probes, we choose the mesonic branch
\beqa
X=\Lambda^{4-4M}\quad ;\quad {\mathcal B}={\widetilde {\mathcal B}}=0 \quad
; \quad \det{\cal M}=\Lambda^{4M}
\eeqa
Restricting to the Abelian case, the superpotential reads
\beqa
W & = & \tr\, \left(- \, C H {\widetilde H}
\, +\, {\widetilde H} H\Phi\, -\, \Phi B\, +\, BC\,\right)
\eeqa
The equation of motion for $B$ requires $\Phi=C$, so we get
\beqa
W&=&- C H {\widetilde H} +  {\widetilde H} H C
\label{n4supo}
\eeqa
The gauge group (in the non-abelian case) is $U(M)$ (due to the breaking 
by meson vevs $M\propto\id$). All three
fields transform in the adjoint representation (a singlet in the abelian 
case). The above theory clearly
describes the field content and superpotential of ${\cal N}=4$ SYM, i.e. 
the theory describing the smooth
geometry left over after the deformation.

In addition, there remain some additional light fields, namely $A$, 
$D$, $B$, subject to the constraint
\beqa
AD-BC=\Lambda^4
\eeqa

The dynamics is that of probe D3-branes in the geometry corresponding to 
the deformation of the SPP to flat space. This matches nicely the 
geometric expectation, from the web diagrams in figure \ref{web_SPP}, from which
we see that the result of the deformation is a smooth geometry.

\medskip

As we have mentioned, for $N=KM$ the field theory is dual to a pure 
supergravity throat with no extra branes (this is achieved by taking the 
baryonic branch in the above analysis of the $SU(M)\times SU(2M)\times 
SU(M)$ theory).
In the main applications in this paper, we are interested in throats with 
a number $p<M$ of D3-brane probes at its tip. Following \cite{ks}, the 
holographic duals of such configurations correspond to the same quiver 
field theory but with $N=KM+p$. In this situation, the next to last step 
of the cascade corresponds to an $SU(p)\times SU(M+p)\times SU(p)$ 
theory. Considering 
the $SU(p)$ theories as global symmetries, we have a $SU(M+p)$ gauge 
theory with $2p$ flavours. Since it has less flavours than colors, the 
theory confines, leaving the mesons (\ref{mesonsspp}) as light degrees of 
freedom, with a non-perturbative Affleck-Dine-Seiberg superpotential 
\cite{ads}
\beqa
W_{ADS}&=&(N_f-N_c)\left(\frac{\Lambda^{3N_c-N_f}}{\det {\cal 
M}}\right)^{\frac{1}{N_c-N_f}}
\eeqa
Following \cite{ks} the superpotential forces $\det{\cal M}$ to acquire a 
non-zero vev, which 
breaks the symmetry down to $SU(p)$. The left over fields are three chiral 
multiplets in the adjoint, described by an $\NN=4$ SYM theory. This is the 
gauge theory on the $p$ D3-branes sitting at the bottom of the throat.
This analysis is particularly simple for $p=1$ (and in particular has 
appeared in \cite{oz}). The complete superpotential reads
\beqa
W & = & AD\,- \, C H {\widetilde H}
\, +\, {\widetilde H} H\Phi\, - \, \Phi B\, +\, (M-1) \left( 
\frac{\Lambda^{3M+1}}{AD-BC}\right)^{\frac{1}{M-1}}
\eeqa
The equation of motion for e.g. $A$ shows that $\det{\cal M}=AD-BC=
\Lambda^{\frac{3M+1}{M}}$. This triggers the promised breaking of 
the symmetry, in which $A$ and $D$ are eaten up. Setting $\Lambda=1$ for 
simplicity in what follows, the equation of motion for $B$ imposes 
$\Phi=C$. Replacing these relations in the above superpotential, the left 
over adjoint fields $H$, ${\widetilde H}$, $C$ have $\NN=1$ superpotential 
interactions, as in (\ref{n4supo}). Notice that although in the abelian 
case it vanishes identically, it is crucial for higher $p$.

\medskip

As mentioned in \cite{fhu}, the relation between the field theory and the 
geometrical description of the deformation can be obtained using the 
construction of the moduli space of the SPP in section \ref{sppmoduli}.
The gauge invariant operators (\ref{mapping}), expressed in terms of the 
mesons, read
\beqa
& x  =  {\widetilde H}D \quad , \quad y  =  AH \quad ,\quad z=\Phi =C 
\quad , \quad w=B ={\widetilde H}H
\eeqa
where we have used F-term relations to express some operator in 
several equivalent ways. The monomials satisfy $xy-zw^2=0$ at the 
classical level, namely
\beqa
{\widetilde H}H \, (\, AD-BC\,)\, =\, 0
\eeqa
On the other hand, at the non-perturbative level we have $AD-BC=\epsilon$.
Hence, the moduli space corresponds to
\beqa
{\widetilde H}H \, (\, AD-BC\,)\, =\, \epsilon {\widetilde H}H 
\eeqa
Namely
\beqa
xy-zw^2=\epsilon w
\eeqa
This is a complex deformation of the SPP to a smooth space.

\subsection{From the parent to the $\IZ_3$ quotient theory}
\label{quot}

{\bf The quotient on the geometry}

As discussed in the main text, the toric data of the geometry of interest indicates that it is a $\IZ_3$ quotient of the SPP. Here we would 
like to be more explicit about this point.

Given a set of toric data for a geometry, namely the charges of a set of  
$p_i$'s under a set of $U(1)$ symmetries, there is a systematic way to 
construct the toric data of a $\IZ_N$ quotient of it. The generator 
$\theta$ of $\IZ_N$ has an action on the $p_i$'s given by a phase rotation 
$e^{2\pi i a_i/N}$ (which is defined modulo the $U(1)$ gauge invariances), 
with $\sum_i a_i=N$. The quotient is described by adding a new variable 
$P$, uncharged under the old gauge symmetries, and a new $U(1)$ 
gauge symmetry, under which the $p_i$ have charges $a_i/N$ and $P$ has 
charge $-1$. The idea is that one can use the new $U(1)$ symmetry 
and its D-term to eliminate $P$. This however still leaves a $\IZ_N$ 
subgroup of the $U(1)$ (phase rotations by $2\pi$) acting 
trivially on $P$ but non-trivially on the $p_i$'s. Modding by this 
discrete gauge symmetry thus implements the orbifold quotient, with 
the desired action. 

We can apply this to show that the quotient of the SPP by the $\IZ_3$ 
action
\beqa
p_5\to \alpha^2 p_5 \quad ; \quad p_6\to \alpha p_6
\eeqa
(modulo gauge equivalences) leads to the toric data of the geometry used 
in the main text. Following the above procedure we add a new field and 
gauge symmetry to (\ref{toriccharges})
\begin{center}
\begin{tabular}{cccccccc}
 & $p_1$ & $p_2$ & $p_3$ & $p_4$ & $p_5$ & $p_6$ & $P$\\
$Q_F$ & $1$ & $-1$ & $1$ & $0$ & $-1$ & $0$ & $0$ \\
$Q_2$ & $-1$ & $1$ & $0$ & $1$ & $0$ & $-1$ & $0$ \\
$Q_3$ & $-1$ & $0$ & $0$ & $-1$ & $1$ & $1$ & $0$ \\ 
$Q_{\IZ_3}$ & $0$ & $0$ & $0$ & $0$ & $2/3$ & $1/3$ & $-1$ 
\end{tabular}
\end{center}
The toric diagram is obtained by computing the cokernel of the charge 
matrix, namely
\beqa
{\widetilde T}  & = &  \pmatrix{ 
0 & 0 & 0 & 3 & 0 & 3 & 1 \cr
1 & 2 & 1 & 2 & 0 & 3 & 1 \cr
1 & 1 & 1 & 1 & 1 & 1 & 1}
\eeqa
which reproduces figure \ref{toricsppz3} up to a linear transformation.

The action on the complex variables $x$, $y$, $z$, $w$ can be obtained by 
use of (\ref{mapping}), and reproduces the one described in the main text 
(\ref{z3action}). Also the action on the fields of the SPP theory 
(\ref{z3actiononfields}) used in the main text is obtained by using 
(\ref{intermsofps}).

{\bf The quotient field theory}

Most results of the main text concerning the quotient of the SPP, and the 
corresponding quiver gauge field theory, can be obtained by considering 
similar effects in the parent SPP field theory, and carrying out the 
$\IZ_3$ quotient \footnote{This is true for any final configuration in 
the quotient of the SPP which respects the $\IZ_3$ quantum symmetry. This 
is not the case e.g. in section \ref{d7branes}, where nodes in the 
quotient arising from a single node in the SPP are given different rank. 
Namely, the model contains fractional D3- and D7-branes with respect to 
the $\IZ_3$ quotient.}. We describe this in the present section.

The construction of the field theory of the quotient of the SPP as a 
quotient of the field theory of the SPP has been described in the main 
text, section (\ref{fieldtheorydual}), and will not be repeated here.

The duality cascade of the quotient theory is simply the quotient of the 
cascade for the SPP theory, studied above. In order to show this it is 
enough to show that Seiberg duality and orbifolding commute. Sketchily, 
one can start with the SPP theory and perform a Seiberg duality on a node. 
Since the $\IZ_3$ is a global symmetry, it exists in both theories, and 
its action of the original and dual fields is related. Carrying out the 
corresponding $\IZ_3$ quotient, one is led to two theories for the 
quotient of the SPP theory. Both quotient theories are related by 
(simultaneous) Seiberg duality on the three nodes corresponding to the 
node dualised in the parent theory. We carry out the exercise in what 
follows.

Consider the field theory for the SPP (\ref{sppfieldth2}). For 
illustration, consider the case with ranks $(N,N+M,N)$, which 
corresponds to the first step in the duality cascade. Carrying out the 
dualization of the node $SU(N+M)$, we obtain the field theory

\begin{center}
\begin{tabular}{cccc}
 & $SU(N)$ & $SU(N-M)$ & $SU(N)$ \\
$F_d$ & $\antifund$ & $\fund$ & $1$ \\
${\widetilde F}_d$ & $\fund$ & $\antifund$ & $1$ \\
$G_d$ & $1$ & $\antifund$ & $\fund$  \\
${\widetilde G}_d$ & $1$ & $\fund$ & $\antifund$  \\
$H$ & $\antifund$ & $1$ & $\fund$  \\
${\widetilde H}$ & $\fund$ & $1$ & $\antifund$  \\
$\Phi'$ & $1$ & $1$ & Adj
\end{tabular}
\end{center}

In the dualization process, we have replaced quarks ${\widetilde F}$, $G$ 
and antiquarks $F$, ${\widetilde G}$ of $SU(N+M)$ by dual quarks $F_d$, 
${\widetilde G}_d$ of $SU(N-M)$, and we have introduced the mesons. However 
the mesons $FG$, ${\widetilde G}{\widetilde F}$ get massive due to the 
superpotential, and the meson $F{\widetilde F}$ becomes massive with the 
original adjoint $\Phi$. The only remaining light meson is 
$\Phi'={\widetilde G}G$.

The $\IZ_3$ action (\ref{z3actiononfields}) on the initial theory is a 
global symmetry, hence a corresponding action is induced on the dual 
theory. The action is obtained by noticing that quarks and dual 
(anti)quarks transform oppositely under global symmetries, and that 
$\Phi'$ is a meson in terms of the original quarks. The action of the dual 
is thus
\beqa
& F_d\to \alpha F_d \quad ; \quad {\widetilde F}_d\to {\widetilde F}_d \nonumber 
\\
& G_d\to G_d \quad ; \quad {\widetilde G}_d\to \alpha^2 {\widetilde G}_d \nonumber 
\\
& H\to \alpha^2 H \quad ; \quad {\widetilde H}\to {\widetilde H} \nonumber \\
& \Phi' \to \alpha \Phi'
\eeqa
It is now possible to carry out the $\IZ_3$ quotient of the dual theory by 
this action. The resulting theory is exactly given by figure 
\ref{dualization}b, namely the theory obtained after carrying out 
simultaneous Seiberg duality of the three $SU(N+M)$ nodes of the $\IZ_3$ 
quotient of the initial SPP theory. This completes our argument that 
orbifolding and Seiberg duality commute. Iteration of the argument leads 
to inheritance of the SPP cascade to the quotient theory.

Finally, it is straightforward to compare the computations of the 
geometric deformation from the field theory, in sections 
(\ref{fieldtheorydual}) and (\ref{sppirdef}), and conclude that the former 
is directly inherited from the latter by a $\IZ_3$ quotient. We leave the 
details to the interested reader.


\begin{thebibliography}{99}
%
\bibitem{rs1}
L. Randall, R. Sundrum, `A Large mass hierarchy from a small extra
dimension', Phys. Rev. Lett. 83 (1999) 3370, hep-ph/9905221.
%
\bibitem{rs2}
L. Randall, R. Sundrum, `An Alternative to compactification',
Phys. Rev. Lett. 83 (1999) 4690, hep-th/9906064.
%
\bibitem{drs}
K. Dasgupta, G. Rajesh, S. Sethi, `M theory, orientifolds and G - flux',
JHEP 9908 (1999) 023, hep-th/9908088.
%
\bibitem{gkp}
S. B. Giddings, S. Kachru, J. Polchinski, `Hierarchies from fluxes in 
string compactifications', Phys. Rev. D66 (2002) 106006, hep-th/0105097.
%
\bibitem{ks}
I. R. Klebanov, M. J. Strassler, `Supergravity and a confining gauge 
theory: Duality cascades and chi SB resolution of naked singularities',
JHEP 0008 (2000) 052, hep-th/0007191.
%
\bibitem{verlinde}
H. Verlinde, `Holography and compactification',
Nucl. Phys. B580 (2000) 264, hep-th/9906182.
%
\bibitem{kw}
I. R. Klebanov, E. Witten, `Superconformal field theory on three-branes at 
a Calabi-Yau singularity', Nucl. Phys. B536 (1998) 199, hep-th/9807080.
%
\bibitem{seiberg}
N. Seiberg, `Electric - magnetic duality in supersymmetric nonAbelian 
gauge theories', Nucl. Phys. B435 (1995) 129, hep-th/9411149.
%
\bibitem{cgqu}
J. F. G. Cascales, M. P. Garcia del Moral, F. Quevedo, A. M. Uranga, 
`Realistic D-brane models on warped throats: Fluxes, hierarchies 
and moduli stabilization', JHEP 0402 (2004) 031, hep-th/0312051.
%
\bibitem{chaotic}
S. Franco, Y.-H. He, C. Herzog, J. Walcher, `Chaotic duality in string 
theory', Phys. Rev. D70 (2004) 046006, hep-th/0402120.
%
\bibitem{ejaz}
Q.J. Ejaz, C.P. Herzog, I.R. Klebanov, `Cascading RG flows from new 
Sasaki-Einstein manifolds', hep-th/0412193.
%
\bibitem{fhu}
S. Franco, A. Hanany, A. M. Uranga, `Multi-flux warped throats
and cascading gauge theories', hep-th/0502113.
%
\bibitem{vafa}
C.~Vafa, `Superstrings and topological strings at large N,'
J.\ Math.\ Phys.\  42 (2001) 2798, hep-th/0008142]; F.~Cachazo, 
K.~A.~Intriligator and C.~Vafa, `A large N duality via a geometric 
transition', Nucl.\ Phys.\ B603 (2001) 3, hep-th/0103067; F.~Cachazo, 
S.~Katz and C.~Vafa, `Geometric transitions and N = 1 quiver theories',
hep-th/0108120; F.~Cachazo, B.~Fiol, K.~A.~Intriligator, S.~Katz and 
C.~Vafa, `A geometric unification of dualities',
Nucl.\ Phys.\ B628 (2002) 3, hep-th/0110028.
%
\bibitem{oda}
I. Oda, `Mass hierarchy from many domain walls', 
Phys. Lett. B480 (2000) 305, hep-th/9908104; `Mass hierarchy and trapping 
of gravity', Phys. Lett. B472 (2000)59, hep-th/9909048.
%
\bibitem{severalRS}
H. Hatanaka, M. Sakamoto, M. Tachibana, K. Takenaga, `Many brane extension 
of the Randall-Sundrum solution', Prog. Theor. Phys 102 (1999) 1213, 
hep-th/9909076.
%
\bibitem{aganagicvafa}
M. Aganagic, C. Vafa, ` G(2) manifolds, mirror symmetry and geometric 
engineering', hep-th/0110171.
%
\bibitem{pqwebs}
O. Aharony, A. Hanany, `Branes, superpotentials and superconformal fixed 
points', Nucl. Phys. B504 (1997) 239, hep-th/9704170;\\
O. Aharony, A. Hanany, B. Kol, `Webs of (p,q) five-branes, 
five-dimensional field theories and grid diagrams',
JHEP 9801 (1998) 002, hep-th/9710116.
%
\bibitem{vafaleung}
N. C. Leung, C. Vafa, `Branes and toric geometry',
Adv. Theor. Math. Phys. 2 (1998) 91, hep-th/9711013.
%
\bibitem{dm}
M.~R.~Douglas, G.~Moore, `D-branes, quivers, and ALE
instantons', hep-th/9603167.
%
\bibitem{dgm}
M. R. Douglas, B. R. Greene, D. R. Morrison, `Orbifold resolution by 
D-branes', Nucl. Phys. B506 (1997) 84, hep-th/9704151.
%
\bibitem{aiqu}
G. Aldazabal, L. E. Ib\'a\~nez, F. Quevedo, A. M. Uranga, `D-branes at 
singularities: A Bottom up approach to the string embedding 
of the standard model', JHEP 0008 (2000) 002, hep-th/0005067.
%
\bibitem{mp}
D. R. Morrison, M. R. Plesser, `Nonspherical horizons. 1',
Adv. Theor. Math. Phys. 3 (1999) 1, hep-th/9810201.
%
\bibitem{uraconi}
A. M. Uranga, `Brane configurations for branes at conifolds',
JHEP 9901 (1999) 022, hep-th/9811004.
%
\bibitem{delpezzo}
B. Feng, A. Hanany, Y.-H. He, `D-brane gauge theories from toric 
singularities and toric duality', Nucl. Phys. B595 (2001) 165
hep-th/0003085;\\
B. Feng, A. Hanany, Y.-H. He, `Phase structure of D-brane gauge theories 
and toric duality', JHEP 0108 (2001) 040, hep-th/0104259;\\
A. Hanany, A. Iqbal, `Quiver theories from D6 branes via mirror symmetry',
JHEP 0204 (2002) 009, hep-th/0108137.
%
\bibitem{ypq}
J. P. Gauntlett, D. Martelli, J. Sparks, D. Waldram, `Sasaki-Einstein 
metrics on $S^2 \times S^3$', hep-th/0403002; D. Martelli, J. Sparks, 
`Toric geometry, Sasaki-Einstein manifolds and a new infinite class of 
AdS/CFT duals', hep-th/0411238;\\
M.~Bertolini, F.~Bigazzi and A.~L.~Cotrone,
  ``New checks and subtleties for AdS/CFT and a-maximization,''
  JHEP {\bf 0412} (2004) 024
  [arXiv:hep-th/0411249];\\
S. Benvenuti, S. Franco, A. Hanany, D. Martelli, J. Sparks, `An Infinite 
family of superconformal quiver gauge theories with Sasaki-Einstein 
duals', hep-th/0411264;\\
S. Benvenuti, A. Hanany, P. Kazakopoulos, `The Toric phases of the 
$Y^{p,q}$ quivers', hep-th/0412279.
%
\bibitem{private}
Private communication from S. Franco, A. Hanany.
%
\bibitem{unhiggs}
B. Feng, S. Franco, A. Hanany, Y.- H. He, `UnHiggsing the del Pezzo',
JHEP 0308 (2003) 058, hep-th/0209228.
%
\bibitem{oz}
H. Ita, H. Nieder, Y. Oz, `On type II strings in two dimensions',
hep-th/0502187.
%
\bibitem{sagn}
A. Sagnotti, `A Note on the Green-Schwarz mechanism in open string 
theories', Phys. Lett. B294 (1992) 196, hep-th/9210127.
%
\bibitem{iru}
L. E. Ib\'a\~nez, R. Rabad\'an, A. M. Uranga, `Anomalous U(1)'s in type I 
and type IIB D = 4, N=1 string vacua', Nucl. Phys. B542 (1999) 112, 
hep-th/9808139.
%
\bibitem{fhhu}
B. Feng, A. Hanany, Y.-H. He, A. M. Uranga, `Toric duality as Seiberg 
duality and brane diamonds', JHEP 0112 (2001) 035, hep-th/0109063.
%
\bibitem{bp}
C. E. Beasley, M. R. Plesser, `Toric duality is Seiberg duality',
JHEP 0112 (2001) 001, hep-th/0109053.
%
\bibitem{ads}
I. Affleck, M. Dine, N. Seiberg, `Dynamical supersymmetry breaking in 
supersymmetric QCD', Nucl. Phys. B241 (1984) 493.
%
\bibitem{kk}
A. Karch, E. Katz, `Adding flavor to AdS / CFT',
JHEP 0206 (2002) 043, hep-th/0205236.
%
\bibitem{cascurcalibra}
J. F. G. Cascales, A. M. Uranga, `Branes on generalized calibrated 
submanifolds', JHEP 0411 (2004) 083, hep-th/0407132.
%
\bibitem{pru}
J. Park, R. Rabad\'an, A. M. Uranga, `Orientifolding the conifold',
Nucl. Phys. B570 (2000) 38, hep-th/9907086.
%
\bibitem{uraquiver}
A. M. Uranga, `From quiver diagrams to particle physics',.
hep-th/0007173.
%
\bibitem{abiu}
R. G. Leigh, M. Rozali, `Brane boxes, anomalies, bending and tadpoles',
Phys. Rev. D59 (1999) 026004, hep-th/9807082; \\
G. Aldazabal, D. Badagnani, Luis E. Ibanez, A. M. Uranga, `Tadpole versus 
anomaly cancellation in D = 4, D = 6 compact IIB orientifolds',
JHEP 9906 (1999) 031, hep-th/9904071;\\
M. Bianchi, J. F. Morales, `Anomalies \& tadpoles',
JHEP 0003 (2000) 030, hep-th/0002149. 
%
\bibitem{ouyang}
P. Ouyang, `Holomorphic D7 branes and flavored N=1 gauge theories',
Nucl. Phys. B699 (2004) 207, hep-th/0311084.
%
\bibitem{strassler}  
M. J. Strassler, `Duality in supersymmetric field theory: General 
conceptual background and an application to real particle physics',
In *Nagoya 1996, Perspectives of strong coupling gauge theories* 237-25. 
%
\bibitem{unification}
L. Randall, M. D. Schwartz, `Quantum field theory and unification in
AdS5', JHEP 0111 (2001) 003, hep-th/0108114; `Unification and the
hierarchy from  AdS5', Phys. Rev. Lett. 88 (2002) 081801,
hep-th/0108115.
%
\bibitem{custodial}
K. Agashe, A. Delgado, M. J. May, R. Sundrum, `RS1, custodial isospin and
precision tests', JHEP 0308 (2003) 050, hep-ph/0308036.
%
\bibitem{kpv}
S. Kachru, J. Pearson, H. Verlinde, `Brane / flux annihilation and the 
string dual of a nonsupersymmetric field theory', JHEP 0206 
(2002) 021, hep-th/0112197.
%
\bibitem{inflation}
G. R. Dvali, Q. Shafi, S. Solganik, `D-brane inflation',
hep-th/0105203;
C.P. Burgess, M. Majumdar, D. Nolte, F. Quevedo, G. Rajesh, R.-J. Zhang, 
`The Inflationary brane anti-brane universe',
JHEP 0107 (2001) 047, hep-th/0105204.
%
\bibitem{kklmmt}
S. Kachru, R. Kallosh, A. Linde, J.Maldacena, L. McAllister, S. P. Trivedi, 
`Towards inflation in string theory', JCAP 0310 (2003) 013,
hep-th/0308055.
%
\bibitem{strings}
E. J. Copeland, R. C. Myers, J. Polchinski, `Cosmic F and D strings',
JHEP 0406 (2004) 013, hep-th/0312067.
%
\bibitem{uralocal}
A.~M.~Uranga, `Local models for intersecting brane worlds',
  JHEP {\bf 0212} (2002) 058
  [arXiv:hep-th/0208014].

\end{thebibliography}
\end{document}